\documentclass{article}

\usepackage{PRIMEarxiv}

\usepackage[utf8]{inputenc} % allow utf-8 input
\usepackage[T1]{fontenc}    % use 8-bit T1 fonts
\usepackage{hyperref}       % hyperlinks
\usepackage{url}            % simple URL typesetting
\usepackage{booktabs}       % professional-quality tables
\usepackage{amsfonts}       % blackboard math symbols
\usepackage{nicefrac}       % compact symbols for 1/2, etc.
\usepackage{lipsum}
\usepackage{fancyhdr}       % header
\usepackage{graphicx}       % graphics
\graphicspath{{media/}}     % organize your images and other figures under media/ folder

\usepackage{verbatim,color,amssymb,epsfig}
\usepackage{graphicx,setspace,float,rotating,longtable,amsmath,amssymb,epsfig,color,lscape,subfigure,url}
\usepackage{epstopdf,multirow,lscape,comment}
\usepackage{multirow}
\usepackage{tabularx}
\usepackage{booktabs}
\usepackage{array,mathabx}
\usepackage[percent]{overpic}
\usepackage{pdfpages}
\usepackage{colortbl}
\usepackage{tikz}
\usepackage{enumitem}
\setlist[itemize]{noitemsep, nolistsep}
\usepackage{algorithm,algorithmic}

\def\refhg{\hangindent=20pt\hangafter=1}

\def\Normal{\hbox{Normal}}

\def\bse{\begin{eqnarray*}}
\def\ese{\end{eqnarray*}}
\def\be{\begin{eqnarray}}
\def\ee{\end{eqnarray}}
\def\bq{\begin{equation}}
\def\eq{\end{equation}}
\def\bse{\begin{eqnarray*}}
\def\ese{\end{eqnarray*}}

\def\wh{\widehat}

\def\b1e{{\mathbf e}}

\def\bw{{\mathbf w}}

\def\bzero{{\mathbf 0}}

\def\bmu{{{\mu}}}
\def\bPhi{{\boldsymbol{\Phi}}}
\def\bbeta{{\boldsymbol{\beta}}}
\newcommand{\bDelta}{\mbox{$\Delta$}}

\def\bA{{\mathbf A}}

\def\bB{{\mathbf B}}
\def\bC{{C}}
\def\b1e{{\mathbf e}}

\def\bB{{\mathbf B}}
\def\br{{\mathbf r}}

\def\bY{{Y}}

\def\bz{{\mathbf z}}

\def\bx{{x}}
\def\bX{{X}}

\def\bU{{\mathbf U}}
\def\bv{{\mathbf v}}

\def\bw{{w}}

\def\bZ{{\mathbf Z}}
\def\bzero{{\mathbf 0}}

\def\bbetakm{{\bbeta^{(k_m)}}}

\def\bbetakmknot{{\bbeta^{(k_m)}_0}}

\newcommand\independent{\protect\mathpalette{\protect\independenT}{\perp}}

\def\independenT#1#2{\mathrel{\rlap{$#1#2$}\mkern2mu{#1#2}}}

\def\bepsilon{{\epsilon}}

\def\Normal{\hbox{Normal}}
\def\Uniform{\hbox{Uniform}}
\def\Bernoulli{\hbox{Bernoulli}}

\newtheorem{Th}{\underline{\bf Theorem}}

\newtheorem{Rem}{\underline{\bf Remark}}

\newtheorem{Cor}{\underline{\bf Corollary}}[Th]

\title{Exploring the validity of the complete case analysis for regression models with a right-censored covariate
}

\author{
  Marissa C. Ashner \\
  Department of Biostatistics \\
  Unversity of North Carolina at Chapel Hill \\
  Chapel Hill, NC 27599\\
  \texttt{mashner@live.unc.edu} \\
   \And
  Tanya P. Garcia \\
  Department of Biostatistics \\
  Unversity of North Carolina at Chapel Hill \\
  Chapel Hill, NC 27599\\
  \texttt{tpgarcia@email.unc.edu} \\
}

\begin{document}
\maketitle

\begin{abstract}
Despite its drawbacks, the complete case analysis is commonly used in regression models with missing covariates. Understanding when implementing complete cases will lead to consistent parameter estimation is vital before use. Here, our aim is to demonstrate when a complete case analysis is appropriate for a nuanced type of missing covariate, the randomly right-censored covariate. Across the censored covariate literature, different assumptions are made to ensure a complete case analysis produces a consistent estimator, which leads to confusion in practice. We make several contributions to dispel this confusion. First, we summarize the language surrounding the assumptions that lead to a consistent complete case estimator. Then, we show a unidirectional hierarchical relationship between these assumptions, which leads us to one sufficient assumption to consider before using a complete case analysis. Lastly, we conduct a simulation study to illustrate the performance of a complete case analysis with a right-censored covariate under different censoring mechanism assumptions, and we demonstrate its use with a Huntington disease data example. 
\end{abstract}

% keywords can be removed
\keywords{complete case \and 
covariate censoring \and missing not at random \and random censoring}

%%%%%%%%%%%%%%%%%%%%%%%%%%%%%%%
\section{Introduction}
\label{sec:introduction}
%%%%%%%%%%%%%%%%%%%%%%%%%%%%%%%

In the ongoing conversation about regression models with missing covariates, complete case analyses, which remove all observations with missingness, almost always have a seat at the table. However, being perceived as the simple, convenient method with many drawbacks, complete case analyses typically take a back seat. Despite its drawbacks, this convenient approach is still used by many analysts, especially newcomers to the field of missing data. Additionally, the complete case analysis is often used as a comparison to novel methods or as an initial estimate for sophisticated approaches. The continual use of the complete case analysis should earn the method a central role in the missing covariate conversation to avoid often overlooked properties and non-unified language surrounding the topic.

Here, we focus on clarifying when and why the complete case analysis leads to a consistent estimator for a nuanced type of missing covariate: right-censored covariates. For a right-censored covariate, the true value of the covariate is unknown due to missingness but is known to be greater than some value. This aspect gives us partial information regarding the unknown truth, which is more than is obtained from a general missing covariate. The body of literature surrounding the statistical handling of censored covariates mentions complete case analyses frequently. However, there are slight variations across the literature in terms of the assumptions made to ensure that the estimator produced by the complete case analysis is consistent. % \cite{RigobonStoker2007, WangFeng2012, Tsimikasetal2012, Atemetal2016,Atemetal2017,Atemetal2017SMMR,Lvetal2017, Kongetal2017, Qianetal2018, MatsouakaAtem2020}. 
Key differences in the assumptions stem from the relationships assumed between the censored covariate, the value at which it is censored, and the censoring indicator. Each source considers one assumption, but assumptions across sources have not been compared before, making it unclear how the assumptions are related, if at all.

Understanding the relationship between the assumptions that drive consistency for a complete case estimator is important for several reasons. First, it is important for more sophisticated procedures for handling censored covariates, like multiple imputation, which uses the partial information available from the censored subject. The consistency of a multiple imputation estimator depends on the correct specification of the ``imputation model'' \cite{Haberetal2021}. For some proposed imputation procedures, the imputation model itself builds from an consistent initial estimator, and in practice, the complete case estimator is often that initial estimator \cite{Bernhardtetal2014, Atemetal2016, Atemetal2017}. This initial use means that if the complete case estimator is not consistent, the imputation model will be misspecified, leading to potential bias in the final estimator. Second, the complete case analysis is the default method for handling any type of missing data in most software packages \cite{Mukakaetal2016}. Many analysts will use these programs without a second thought, so we are advocating for further awareness of when the default is valid and how to ensure the censored data are set up properly to handle the default. Finally, when new methods are proposed, many statisticians use simulations to compare the performance of their method to others, including the complete case analysis, which serves as a useful baseline (e.g., see \cite{Bernhardt2018, MatsouakaAtem2020}). It is crucial to be aware of the properties of \textit{all} methods considered in simulations to fully understand the results.

In this paper, we will lay a clear foundation for what assumptions drive consistency when using a complete case analysis for regression models with a censored covariate. We clarify that, in fact, the different assumptions are interconnected in a hierarchical fashion and share a common property: either the censoring value or the censoring indicator must be independent of the outcome, given the covariates. The rest of the paper is organized as follows: Section \ref{sec:censoring} details and condenses the censoring mechanism assumptions present in the existing literature and proves the relationship between the mechanism assumptions as well as the consistency of the complete case estimator under said assumptions; Section \ref{sec:simulations} contains simulation studies conducted under many settings to compare a complete case analysis to the analysis that would be done if there were no censoring; Section \ref{sec:realdata} applies the complete case analysis to a Huntington disease observational study with a censored covariate; and Section \ref{sec:discussion} presents a discussion of our findings. 

%%%%%%%%%%%%%%%%%%%%%%%%%%%%%%%
\section{Establishing consistency of the complete case estimator}
\label{sec:censoring}
%%%%%%%%%%%%%%%%%%%%%%%%%%%%%%%

%%%%%%%%%%%%%%%%%%%%%%%%%%%%%%%
\subsection{The complete case estimator with a right-censored covariate}
\label{sec:notation}
%%%%%%%%%%%%%%%%%%%%%%%%%%%%%%%

We consider the regression model \be
\label{eqn:equation}
{Y}_i = m(X_i, \bZ_{i}, \bbeta) + {\epsilon}_i, \qquad i = 1, \ldots, n, \ee
where $Y_i$ is the outcome, $X_i$ is a right-censored covariate, $\bZ_i$
is a $p$-dimensional vector of fully observed covariates, $\bepsilon_i$ is random error, and $n$ is the number of observations. We will use the subscript $i$ only when needed for clarity. Rather than assume a particular distributional assumption for the random error, we assume that $E(\epsilon|X,\bZ) = 0$ to make our regression model flexible enough to handle a range of random error distributions. In equation \eqref{eqn:equation}, $m(\cdot)$ is any linear or nonlinear function known up to $\bbeta$, a $(p+1)$-dimensional parameter. Due to right-censoring, rather than observe $X$, we observe $W=\min(X,C)$ and $\Delta=I(X\leq C)$, where $C$ is a random variable equal to the value at which $X$ is censored.

One of our goals is to show when the complete case estimator for $\bbeta$ is consistent; we first show how to construct the complete case estimator. The estimators in this paper will be solutions to estimating equations. For a full dataset (i.e., $X$ is not censored), the estimator for $\bbeta$ is the solution to the estimating equation
\be
\label{eqn:oracle-estimating-equation}
\frac{1}{n}\sum_{i=1}^n \bPhi(Y_i, X_i, \bZ_i;\bbeta) = \bzero,
\ee
where $\bPhi(Y,X,\bZ;\bbeta)=\bA(X,\bZ;\bbeta)\{Y-m(X,\bZ;\bbeta)\}$ and $\bA(X,\bZ;\bbeta)$ is any $(p+1)$ dimensional function. The optimal choice of $\bA(X,\bZ;\bbeta)$, in terms of maximal efficiency, is when $\bA(X,\bZ;\bbeta) = \partial m(X,\bZ;\bbeta)/\partial \bbeta \times E(\epsilon^2|X,\bZ)^{-1}$ \cite{Tsiatis2006}. In this paper, we will assume the error term is independent of the covariates such that $E(\epsilon^2|X,\bZ) = E(\epsilon^2)$, making $\bA(X,\bZ;\bbeta) = \partial m(X,\bZ;\bbeta)/\partial \bbeta$ optimal.

The complete case analysis discards all data for which $X$ is right-censored. Mathematically, the complete case estimator, denoted by $\wh\bbeta$, is the solution to
\be
\label{eqn:complete-case-estimating-equation}
\frac{1}{n}\sum_{i=1}^n \Delta_i\bPhi(Y_i, W_i, \bZ_i;\bbeta) = \bzero,
\ee
where we add $\bDelta_i$ so that there is no contribution from each observation with a censored $X_i$. Even though, in general, there is no closed form for $\wh\bbeta$, it can be solved using numerical techniques in available software packages, such as {\tt nls} in R; we provide working code at [URL redacted for blinded submission]. %\href{https://github.com/marissaashner/Complete_Case_Censored_Covariates}{https://github.com/marissaashner/Complete\_Case\_Censored\_Covariates}.
Note that the default for packages like {\tt nls} is to delete observations with a \textit{missing} covariate, so before using these packages on censored data, we must ensure that the censored $X$ values are coded as missing values, not as the value at which it was censored. To show consistency of the complete case estimator, $\wh\bbeta$, additional assumptions are needed. To set the stage, we first draw connections to missing data assumptions. 

%%%%%%%%%%%%%%%%%%%%%%%%%%%%%%
\subsection{Connections to missing data}
\label{sec:missingdata}
%%%%%%%%%%%%%%%%%%%%%%%%%%%%%%

At first glance, a right-censored covariate seems no different than a missing covariate. Yet, even though censoring is a type of missingness, it is different. With general missingness, we have no information about the true covariate value. With censoring, though, we have \emph{partial information}: if $X$ is right-censored, its true value must be larger than $C$. Statistical methods for handling missing data do not adjust for such partial information \cite{Little1992}. 

That said, methods and language used in the missing data context can serve as a starting point for handling right-censored covariates. For example, when dealing with missing data, it is fundamental to consider the so-called missingness mechanisms \textendash assumptions that explore the relationship between the missingness variable (i.e., $\Delta$) and the data values (i.e., $X, Y, \bZ$) \textendash to explain why missing data occur \cite{Little2002, BaraldiEnders2010}. The three most common missingness mechanisms are: (1) missing completely at random, when the distribution of the missing data indicator is independent of all other variables; (2) missing at random, when the distribution of the missing data indicator depends on fully observed variables; and (3) missing not at random, when the distribution of the missing data indicator depends on the missing values themselves.

Models with a right-censored covariate case fall into the category of missing not at random because the ``missing data" indicator is $\bDelta = I(\bX \leq \bC)$, which explicitly depends on incomplete $\bX$. Data that are missing not at random must be considered carefully, since many existing methods require missing at random data \cite{WangFeng2012}. However, an underappreciated fact is that a complete case analysis can lead to consistent estimators for $\bbeta$ even when data are missing not at random \cite{Bartlettetal2014}. This consistency holds so long as certain assumptions about the missingness mechanisms hold. From now on, we will refer to missingness mechanisms as ``censoring mechanisms'' in the context of censored covariates, and we clarify the censoring mechanisms that lead to consistent complete case estimators next. 

%%%%%%%%%%%%%%%%%%%%%%%%%%%%%%%
\subsection{Censoring mechanisms for consistent complete case estimators}
\label{sec:mechanisms}
%%%%%%%%%%%%%%%%%%%%%%%%%%%%%%%

Unlike missingness mechanisms for missing data, the language surrounding censoring mechanisms for censored covariates is not yet unified. Literature for censored covariate problems use different censoring mechanism assumptions to ensure a complete case estimator is consistent, and it is unclear which censoring mechanism an analyst working on a new problem should consider. We aim to unify this language by first summarizing five common censoring mechanisms that current sources use. Those censoring mechanisms are as follows, where the symbol $\independent$ denotes independence between random variables:

\begin{enumerate}[label=(C\arabic*)]
  \item\label{assump:exogenous} \textbf{Exogenous Censoring: } $E(\epsilon|\Delta, X, \bZ) = 0$ \cite{RigobonStoker2007, WangFeng2012, Tsimikasetal2012};
  
  \item\label{assump:strictexogenous} \textbf{Strict Exogenous Censoring: } $\bepsilon \independent \bDelta|(\bX, \bZ)$ \cite{RigobonStoker2007};
  
   \item\label{assump:condxz} \textbf{Conditionally Independent Censoring given $(\bX,\bZ)$: } $\bC \independent \bY | (\bX, \bZ)$ \cite{Atemetal2016, Atemetal2017, Lvetal2017, Qianetal2018};
   
  \item\label{assump:condz} \textbf{Conditionally Independent Censoring given $\bZ$: } $\bC\independent (\bX, \bY)|\bZ$ \cite{Atemetal2017SMMR, Kongetal2017};
   
   \item\label{assump:indep} \textbf{Independent Censoring: } $\bC\independent(\bY, \bX, \bZ)$ \cite{MatsouakaAtem2020}.

\end{enumerate}

%\vspace{5mm}

All five mechanisms have one common feature: either the censoring variable $\bC$ or the censoring indicator $\bDelta$ is independent of the outcome $\bY$ or error term $\epsilon$ in some way. There are still, however, differences between these mechanisms. 
As we move down the list of censoring mechanisms from \ref{assump:exogenous} to \ref{assump:indep}, the statements appear to get more restrictive in that each censoring mechanism imposes a more predetermined structure than the last.

% Comparing 1 and 2 
Recall our model assumption that $E(\epsilon|X, \bZ) = 0$. Censoring mechanism \ref{assump:exogenous} imposes a stronger assumption on the model error such that, rather than requiring the conditional mean of $\epsilon$ to equal 0, it requires the conditional mean of $\epsilon$ among those where $\Delta = 0$ to be 0 \textit{and} the conditional mean of $\epsilon$ among those where $\Delta = 1$ to be 0. Mechanism \ref{assump:strictexogenous} further requires that the conditional distributions of $\bepsilon$ and $\bDelta$ are independent given $X$ and $\bZ$. This means that not only is the conditional mean of $\bepsilon$ independent of $\Delta$ but also the entire probability distribution of $\bepsilon$ is independent of $\Delta$. Censoring mechanism \ref{assump:condxz} switches things up by considering the distribution of the censoring variable $\bC$, rather than the censoring indicator $\bDelta$. Under mechanism \ref{assump:strictexogenous}, we expect that the value of \textit{binary} variable $\Delta$ is independent of the conditional density of $\bepsilon$, and under mechanism \ref{assump:condxz}, we expect that the value of \textit{continuous} variable $C$ is independent of the conditional probability density of $\bY$. Censoring mechanisms \ref{assump:condz} and \ref{assump:indep} are very similar to mechanism \ref{assump:condxz} in that they require independence of censoring variable $\bC$ and outcome $\bY$, but differ in that they condition on fewer variables. 

At face value, it is unclear if these censoring mechanisms are formally related to one another. Will the complete case estimator be consistent if we assume that just one of the assumptions holds, or must we consider them all? In fact, we
show in Theorem \ref{thm:censoring-relationships} that there is indeed a hierarchical relationship between censoring mechanisms \ref{assump:exogenous}\textendash\ref{assump:indep}.

\begin{Th}
\label{thm:censoring-relationships} 
Consider the setup in Section \ref{sec:notation} and the censoring mechanisms \ref{assump:exogenous}\textendash\ref{assump:indep}. Then the following relationship holds between the mechanisms: 
Censoring mechanism \ref{assump:indep} implies \ref{assump:condz}, \ref{assump:condz} implies \ref{assump:condxz}, \ref{assump:condxz} implies \ref{assump:strictexogenous}, and \ref{assump:strictexogenous} implies \ref{assump:exogenous}. Given this hierarchical relationship, we have the following contrapositive: if censoring mechanism \ref{assump:exogenous} does not hold, then \ref{assump:strictexogenous} does not hold; if \ref{assump:strictexogenous} does not hold, \ref{assump:condxz} does not hold; if \ref{assump:condxz} does not hold, \ref{assump:condz} does not hold; and if \ref{assump:condz} does not hold, \ref{assump:indep} does not hold.
\end{Th}

%\section{Using the Censoring Mechanisms in Practice}
%\label{sec:practice}

%\subsection{The hierarchical relationship between the censoring mechanisms}
%\label{sec:relationships}

%%%%%%%%%%%%%%%%%%%%%%%%%%%%%%%%%%%%%%%%%%%%
%\subsection{Consistency of the complete case estimator}
%\label{sec:consistency}
%%%%%%%%%%%%%%%%%%%%%%%%%%%%%%%%%%%%%%%%%%%%

The proof of Theorem \ref{thm:censoring-relationships} is in 
 Supplementary Materials \ref{app:th1}. This hierarchy helps to establish which censoring mechanism is sufficient to ensure the complete case estimator is consistent and asymptotically normal, as described in Theorem \ref{thm:exogenous-consistent}.

\begin{Th}
\label{thm:exogenous-consistent} 
Under the following regularity conditions
\begin{enumerate}[label=(\emph{R\arabic*})]
  \item \label{reg:unbias}
  $E\{\bPhi(\bY, \bX, \bZ;\bbeta)\} = \bzero$ for all $\bbeta$;
  \item \label{reg:nonsing}
  $E\{\frac{\partial\bPhi(\bY, \bX, \bZ;\bbeta)}{\partial \bbeta^T}\}$ is nonsingular;
  \item \label{reg:uniform}
  $n^{-1} \sum_{i=1}^n \frac{\partial\bPhi_i(\bY, \bX, \bZ;\bbeta_0)}{\partial \bbeta^T}$ converges uniformly to $E\{\frac{\partial\bPhi(\bY, \bX, \bZ;\bbeta_0)}{\partial \bbeta^T}\}$ in a neighborhood of $\bbeta_0$, where $\bbeta_0$ is the true parameter value,
\end{enumerate}
the complete case estimator $\wh\bbeta$ is consistent if $E(\epsilon|X,\Delta,\bZ)=0$ (i.e., we have exogenous censoring); that is, $\wh{\bbeta}$ converges in probability to the true parameter value $\bbeta_0$. Moreover, the asymptotic distribution of $\wh{\bbeta}$ will be such that 
$$\sqrt{n}(\wh{\bbeta}-\bbeta_0) \rightarrow \Normal\{\bzero, \bA^{-1}\bB(\bA^{-1})^T\},$$
where $\bA = E\{\partial \Delta\bPhi(Y,X,\bZ;\bbeta_0)/\partial\bbeta^T\}$ and $\bB = E[\{\Delta\bPhi(Y,X,\bZ;\bbeta_0)\}\{\Delta\bPhi(Y,X,\bZ;\bbeta_0)\}^T]$.
\end{Th}
 The proof of Theorem \ref{thm:exogenous-consistent} is in Supplementary Materials \ref{app:consistent_cc_nonlinear}. 
 Theorem \ref{thm:exogenous-consistent} establishes consistency when censoring mechanism \ref{assump:exogenous} holds. Corollary \ref{theorem:corollary} will establish that the complete case estimator is consistent under \emph{any} one of the censoring mechanisms \ref{assump:exogenous} to \ref{assump:indep}.

\begin{Cor}
\label{theorem:corollary} Under any of the censoring mechanisms \ref{assump:exogenous} to \ref{assump:indep}, the complete case estimator $\wh\bbeta$ %(i.e., the solution to equation \eqref{eqn:complete-case-estimating-equation}) 
is a consistent and asymptotically normal estimator. 
\end{Cor}
Corollary \ref{theorem:corollary} follows directly from Theorems \ref{thm:censoring-relationships} and \ref{thm:exogenous-consistent}. The hierarchy in Theorem \ref{thm:censoring-relationships} shows that if any one of the censoring mechanisms \ref{assump:strictexogenous} to \ref{assump:indep} hold, then \ref{assump:exogenous} holds. From Theorem \ref{thm:exogenous-consistent}, if mechanism \ref{assump:exogenous} holds, then the estimator $\wh\bbeta$ is consistent. Therefore, combining the results from Theorems \ref{thm:censoring-relationships} and \ref{thm:exogenous-consistent} leads to the result in Corollary \ref{theorem:corollary}: any of the five censoring mechanisms will lead to a consistent estimator for $\bbeta$.

%%%%%%%%%%%%%%%%%%%%%%%%
\subsection{Unbiasedness of the linear complete case estimator}
\label{sec:bias_linear}
%%%%%%%%%%%%%%%%%%%%%%%

%Now that we have proven consistency properties for a general mean function, we explore some specific results that occur when using a linear mean function. 

Consistency is a desirable property for any estimator, since it ensures that $\wh\bbeta$ converges to the true parameter value, $\bbeta_0$, in probability. Convergence in probability is a large sample property meaning that the estimator will have higher probability of being ``close'' to the true value as the sample size increases to infinity. But what happens when we have a finite sample size? At what point will it be large enough for the large sample property to start ``working''? We show that when the regression model in equation \eqref{eqn:equation} is linear, then the complete case estimator exhibits unbiasedness, which is a finite sample property.

\begin{Rem}
\label{theorem:linear_unbias}
Consider a linear regression model with a right-censored covariate where in equation \eqref{eqn:equation}, $m(X,\bZ;\bbeta) = \beta_0 + \beta_1 X + \bbeta_2^T\bZ$. Let
$\bU_i = \begin{bmatrix} 1 & X_i & \bZ_i^T \end{bmatrix}$ be the $i$th row of the design matrix associated with this linear regression model. If and only if exogenous censoring \ref{assump:exogenous} holds, then the complete case estimator found by solving Equation \eqref{eqn:complete-case-estimating-equation}, which is the least squares solution $\wh{\bbeta} = (\sum_{i=1}^n \Delta_i\bU_i^T\bU_i)^{-1}\sum_{i=1}^n \Delta_i\bU_i^TY_i$, is not only a consistent estimator but is also unbiased (i.e., $E(\wh{\bbeta}) = \bbeta_0$). 
\end{Rem}

Remark \ref{theorem:linear_unbias} says that, for any finite sample, the expectation of the estimator $\wh{\bbeta}$ will equal the true parameter $\bbeta_0$. Similar to Corollary \ref{theorem:corollary} for consistency, we have unbiasedness when any of the censoring mechanisms \ref{assump:exogenous}\textendash\ref{assump:indep} hold. The proof of unbiasedness under exogenous censoring is similar to that used for linear regression models when no covariate is right-censored; see for example \cite{RencherSchaalje2008}. 

Having this result allows us to investigate the bias that can occur when exogenous censoring \ref{assump:exogenous} does not hold. In Supplementary Materials \ref{app:bias_linear_int}, we show that for a linear model with an intercept, only the parameter estimate for the intercept (i.e., $\wh{\beta}_0$) will be biased if exogenous censoring does not hold, and the other estimates remain unbiased. Additionally, if the intercept is removed from the model, as in $m(\bX,\bZ;\bbeta) = \beta_1X + \bbeta_2\bZ$, it is possible to see bias in the parameter estimates for $\beta_1$ and $\bbeta_2$.

%%%%%%%%%%%%%%%%%%%%%%%%%%%%%%%%%%%%%%%%%%%%
\subsection{Checking the censoring mechanism assumptions}
\label{sec:check-assump}
%%%%%%%%%%%%%%%%%%%%%%%%%%%%%%%%%%%%%%%%%%%%

While censoring mechanisms \ref{assump:exogenous}\textendash\ref{assump:indep} help to ensure consistency of the complete case estimator, how can an analyst check when these censoring mechanisms hold? Unfortunately, verifying any assumptions made on the censoring mechanisms is nearly impossible based on the observed data alone, since they all rely on incomplete $\bX$ \cite{WhiteCarlin2010}. 

However, we may still be able to gauge from contextual knowledge of the data whether or not these censoring mechanisms hold. One way to visually explore the assumptions about independence between variables is to use a directed acyclic graph (DAG) \cite{Bartlettetal2014}. DAGs are commonly used in the context of causal inference and are a tool used to visualize our contextual knowledge and \textit{a priori} assumptions about the causal relationships between the variables in our model. DAGs consist of nodes representing random variables and edges representing the relationships between these variables. The edges have arrows, meaning they imply a direction; an arrow from $\bX$ to $\bY$ implies that $\bX$ causes $\bY$. After using this framework to lay out the hypothesized causal structure of the variables of interest, the DAG can be used to describe marginal and conditional independence between variables. To do so, let us define a few terms: 
\begin{enumerate}
  \item A \textbf{path} between two variables on a DAG is a route that connects the two variables by following a sequence of edges;
  \item A \textbf{collider} on a path is a node on the path where two arrowheads collide.
\end{enumerate}
Two variables are marginally independent if, for all paths between the two, there is at least one collider. Two variables are conditionally independent if, for all paths between the two, there is at least one collider and/or we condition on at least one other variable (a non-collider) on the path \cite{HernanRobins2020}. 

This concept can be best understood through an example. Consider a longitudinal observational study of patients who tested positive for the genetic mutation that causes Huntington disease. Let $\bX$ be the time to clinical diagnosis (in years, which may be right-censored). $\bDelta$ is the indicator of whether or not $\bX$ is censored, and $\bC$ is the time to censoring (in years). Let us define $\bZ$ as the number of cytosine, adenine, guanine (CAG) repeats in the patient's huntingtin gene and their current age, which are both factors known to be associated with impairment in this disease \cite{Zhangetal2011}. We will look at two outcomes $\bY$. First, let $\bY$ be the current level of apathy felt by the patient. Then, let $\bY$ be the current motor impairment in Huntington disease patients (called the total motor score), where a higher score means greater impairment. We hypothesize that some people dropped out of the study because their motor impairment prevented them from continuing. By this we mean, a patient's motor impairment could eventually limit their physical ability to continue participating in the study.

\begin{figure}[htb!]
\centering
\subfigure[]{%
\includegraphics[width=0.45\textwidth]{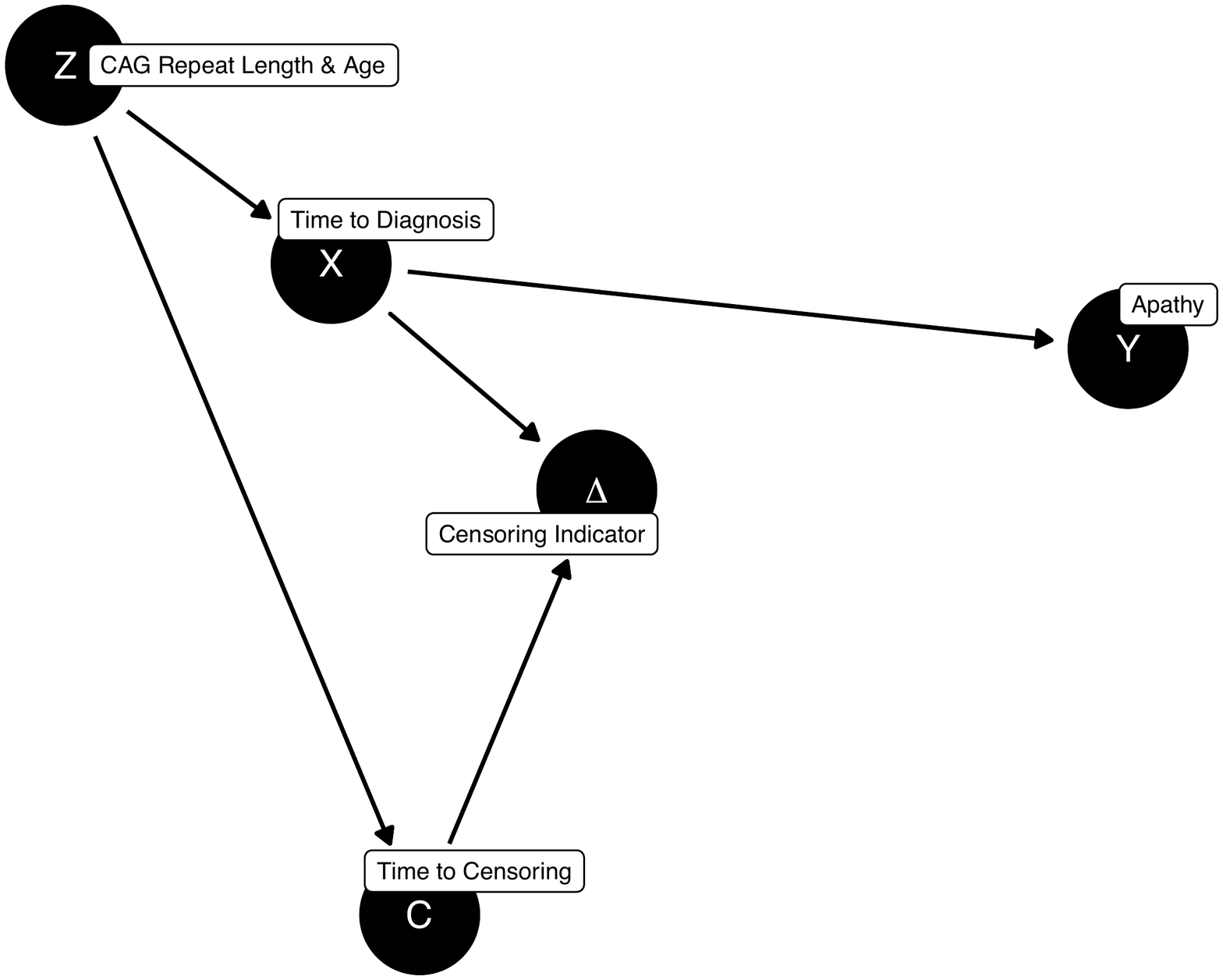}
\label{fig:cc_dag_a}%
}%
\hskip 0.2in
\subfigure[]{%
\includegraphics[width=0.45\textwidth]{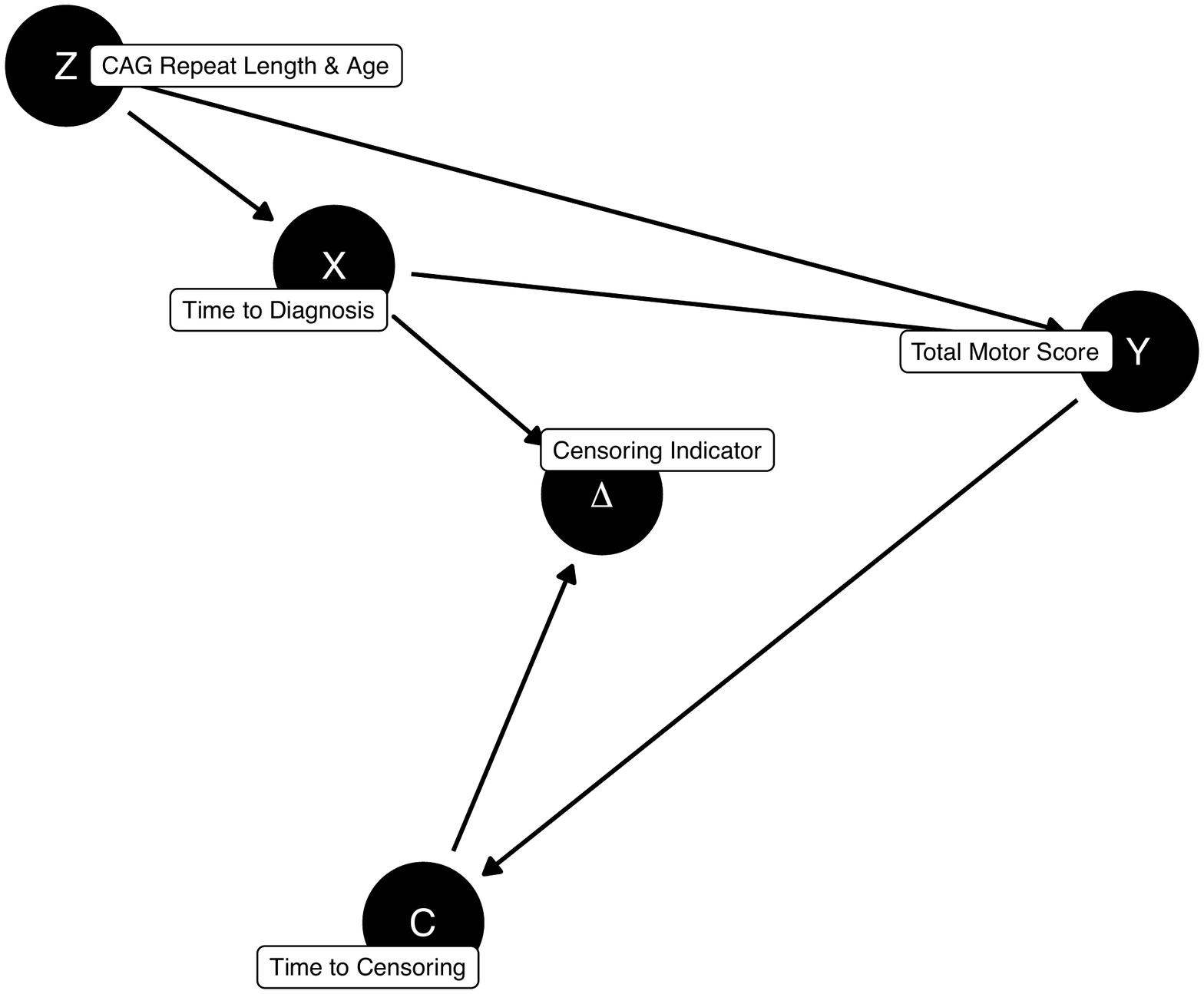}
\label{fig:cc_dag_b}
}%
\caption{
Directed acyclic graphs (DAGs) that represent the hypothesized causal relationships in the Huntington disease example for an apathy outcome (a) and a total motor score outcome (b).
}
\label{fig:cc_dag}
\end{figure}

First consider the DAG in Figure \ref{fig:cc_dag_a}. We draw a directed edge from $\bX$ to $\bY$ since this is our primary association of interest, and directed edges from $\bX$ and $\bC$ to $\bDelta$, since by definition, the censoring indicator is associated with both $\bX$ and $\bC$. Now, we draw a directed edge from $\bZ$ to $\bX$, since it has been shown in previous studies that CAG repeat length and age are predictive of disease diagnosis \cite{Zhangetal2011}. We also draw an edge from $\bZ$ to $\bC$, since one could argue that CAG repeat length and age may lead to an earlier dropout . We want to determine whether or not a complete case analysis is consistent based on this DAG. Consider the following paths from $\bC$ to $\bY$: (1) $\bY \leftarrow \bX \leftarrow \bZ \rightarrow \bC$; and (2) $\bY \leftarrow \bX \rightarrow \bDelta \leftarrow \bC$. If we condition on $\bX$ and $\bZ$, then all paths between $\bC$ and $\bY$ meet the criteria for conditional independence, so $\bY \independent \bC | (\bX, \bZ)$ and censoring mechanism \ref{assump:condxz} holds. This means we have consistency of the complete case estimator. 

Now, consider the DAG in Figure \ref{fig:cc_dag_b}. Most of the directed edges are hypothesized to match Figure \ref{fig:cc_dag_a}, but one stark distinction is the directed edge from $\bY$ to $\bC$. We might hypothesize that the current total motor score is directly associated with the time to censoring, since patients with a higher total motor score may drop out sooner. We cannot condition on any variables in this direct path, so $\bC$ is dependent on $\bY$. We therefore think the censoring mechanisms may not hold, meaning we may not have guaranteed consistency using a complete case estimator. 

It is important to recognize that these conclusions are not formal, since the DAGs are made from \textit{a priori} assumptions regarding subject-matter knowledge. One could instead argue that in Figure \ref{fig:cc_dag_a}, apathy is associated directly with the time to censoring, since apathetic people may lose interest in the study. We also note that we cannot consider exogenous censoring from DAGs, which relies on an expectation assumption rather than an independence assumption. However, DAGs are still helpful for visualizing the mechanisms.

%%%%%%%%%%%%%%%%%%%%%%%%%%%%%%%
\section{Simulation Study}
\label{sec:simulations}
%%%%%%%%%%%%%%%%%%%%%%%%%%%%%%%

 We conduct a simulation study to empirically demonstrate the consistency of the complete case estimator. Namely, we generate data under six different combinations of censoring mechanism assumptions to show that a complete case estimator is consistent when at least mechanism \ref{assump:exogenous}, i.e., exogenous censoring, holds. 

\subsection{Data Generation and Metrics for Comparison}
\label{sec:sims_data}

Our simulations use sample sizes $n=400$ and $n=1200$ and censoring rates of $25\%$ and $75\%$. The combination of $n=1200$ and a censoring rate of $75\%$ is a similar dataset size to that of the real data example that will be shown in Section \ref{sec:realdata}. We also use a smaller sample size and smaller censoring rate to illustrate other dataset types seen in practice.

We generate data for two regression models in equation \eqref{eqn:equation}: a linear model where $m(X,Z;\bbeta) =0.5 + X - 2Z$ and a nonlinear logistic model where $m(X,Z;\bbeta) = 1/[1+\exp\left\{-5(0.005 + 0.01X -0.02Z)\right\}]$. One thousand replications are generated for each combination of sample size, censoring rate, and mean function.

For all settings, we generate $\bX \sim \Uniform(0,3)$ and $Z \sim \Normal(0,1)$. The generation of $\bC$, $\bDelta$, and $\bepsilon$ vary depending on which of the censoring mechanisms hold. The choices made for the data generation of $\bC$, $\bDelta$, and $\bepsilon$ are described in Table \ref{tab:sim_settings}, and the justification for these data generation choices are described in Supplementary Materials \ref{app:sims}.

\begin{table}
\centering
\caption{Data generation for six simulation settings, each with a different combination of censoring mechanism assumptions. For some settings, $\Delta$ is generated directly and $C$ is left unspecified, and for the rest, $C$ is generated directly and $\Delta = I(X\leq C)$. Let $r$ be the censoring rate in decimal form.}
\resizebox{\linewidth}{!}{
\begin{tabular}[t]{ >{\centering\arraybackslash}p{5cm} >{\centering\arraybackslash}p{6.6cm} >{\centering\arraybackslash}p{6.1cm}}
 \toprule
% & & \multicolumn{2}{c}{Bias (Empirical Standard Error)} \\
 \textbf{Censoring Mechanism Setting} & \textbf{Censoring Variable $\bDelta$ or $\bC$} & \textbf{Error Term $\bepsilon$} \\
 \midrule
 % Setting 1
\textbf{Exogenous Censoring does not hold:} \ref{assump:exogenous}\textendash\ref{assump:indep} do not hold & $\bDelta \sim \Bernoulli(1-r)$ & $\Normal(\mu, \sigma^2)^{1}$ where $\mu = \frac{\sigma}{8}(1-\bDelta) - \frac{\sigma}{8}\frac{r}{1-r}\bDelta$\\
 
 \midrule
 % Setting 2
\textbf{Strict Exogenous Censoring does not hold:} \ref{assump:exogenous} holds and \ref{assump:strictexogenous}\textendash\ref{assump:indep} do not hold & $\bDelta = \begin{cases}
1, & |\bepsilon| \geq \text{the rth quantile of } |\bepsilon|, \\ 0 & \text{otherwise} \end{cases}$ & $\Normal(0,\sigma^2)$\\

\midrule
% Setting 3
\textbf{Conditional Independence given (X,Z) does not hold:} \ref{assump:exogenous}\textendash\ref{assump:strictexogenous} hold and \ref{assump:condxz}\textendash\ref{assump:indep} do not hold & $\bC \sim \begin{cases} \Uniform(0, 6-6r) & r \geq 0.5,\\ \Uniform(0, \frac{3}{2r}) & \text{otherwise} \end{cases}$ & $\bepsilon \sim \begin{cases} \Normal(0, 0.5\sigma^2) & C \in \\ & \text{IQR}(\bC)^{2}, \\ \Normal(0, 1.5\sigma^2) & \text{otherwise} \end{cases}$ \\

\midrule
% Setting 4 
\textbf{Conditional Independence given Z does not hold:}  \ref{assump:exogenous}\textendash\ref{assump:condxz} hold and \ref{assump:condz}\textendash\ref{assump:indep} do not hold & $\bC \sim \Uniform(0, X/r)$ & $\Normal(0,\sigma^2)$ \\

\midrule
% Setting 5 
\textbf{Independence does not hold:} \ref{assump:exogenous}\textendash\ref{assump:condz} hold and \ref{assump:indep} does not hold & $\bC \sim \begin{cases} \Uniform(0, 6-6r) & r\geq 0.5 \& \\ & Z > \text{med}(Z_1\ldots,Z_n)^{3}, \\ \Uniform(\frac{3-6r}{2-2r}, 3) & r\geq 0.5\& \\ & Z \leq \text{med}(Z_1\ldots,Z_n), \\ \Uniform(0, \frac{3}{2r}) & r< 0.5\& \\ & Z > \text{med}(Z_1\ldots,Z_n), \\ \Uniform(3-6r, 3) & r< 0.5\& \\ & Z \leq \text{med}(Z_1\ldots,Z_n) \end{cases}$ & $\Normal(0,\sigma^2)$\\

\midrule
% Setting 6 
\textbf{Independence holds:} \ref{assump:exogenous}\textendash\ref{assump:indep} hold & $\bC \sim \begin{cases} \Uniform(0, 6-6r) & r \geq 0.5, \\ \Uniform(0, \frac{3}{2r}) & \text{otherwise} \end{cases}$ & $\Normal(0,\sigma^2)$ \\

  \bottomrule
  \multicolumn{3}{l}{$^{1}$\footnotesize{Note that $\sigma^2 = 2$ for the linear mean model and $\sigma^2 = 0.08$ for the logistic model.}}\\
  \multicolumn{3}{l}{$^{2}$\footnotesize{IQR$(\cdot)$ represents the interquartile range.}}\\
  \multicolumn{3}{l}{$^{3}$\footnotesize{med$(\cdot)$ represents the median.}} \\
\end{tabular}}
\label{tab:sim_settings}
\end{table}

For each setting, we estimate the regression parameters $\bbeta$ using two methods. First, we use a complete case estimator, where only observations with an observed $\bX$ were included in the final sample. Secondly, we use a so-called oracle estimator, where all observations are kept in the analysis, rather than deleting those that are censored. This method is not possible in practice since $\bX$ is censored, but it gives us a gold standard to compare the complete case analysis to. We compare results from the complete case and oracle estimators in terms of empirical bias and estimated variability. We report bias in terms of percent bias calculated as $1/1000\sum_{i=1}^{1000}(\wh{\bbeta}_i-\bbeta)/\bbeta\times 100,$ and the estimated variability in terms of the empirical mean of the estimated standard error calculated as $1/1000\sum_{i=1}^{1000} \wh{SE}_i$, where $\wh{SE}_i$ is calculated from the sandwich variance estimator. The sandwich estimator is an estimate of the asymptotic variance of $\wh{\bbeta}$ outlined in Theorem \ref{thm:exogenous-consistent}. Namely, $\wh{Var}(\wh{\bbeta}) = \wh{\bA}^{-1}\wh{\bB}(\wh{\bA}^{-1})^T$
where $\wh{\bA} = n^{-1}\sum_{i=1}^n \partial\Delta_i\bPhi(Y_i,X_i,\bZ_i;\wh{\bbeta})/\partial\bbeta^T$, $\wh{\bB} = n^{-1}\sum_{i=1}^n \{\Delta_i\bPhi(Y_i,X_i,\bZ_i;\wh{\bbeta})\}\{\Delta_i\bPhi(Y_i,X_i,\bZ_i;\wh{\bbeta})\}^T$, and $\wh{SE} = \sqrt{\wh{Var}(\wh{\bbeta})/n}$. Additionally, we calculate the observed coverage probability for a Wald-type 95\% confidence interval for each method. Since we are using the oracle estimator as a gold standard, we want to determine under what simulation settings (i.e., considering sample size, censoring rate, mean function, and censoring mechanisms) these metrics are comparable between the two methods and under what settings they are substantially different.

\subsection{Results}
\label{sec:sims_results}

\begin{table}
\caption{\label{tab:sims_400_linear} Simulation results for a linear model with a right-censored covariate when data are generated under six different combinations of censoring mechanism assumptions that correspond with those in Table \ref{tab:sim_settings}. We report the percent bias, estimated standard errors (SE), and 95\% coverage probabilities for all parameters in the linear model when we estimate parameters using the complete case estimator (CC) and oracle estimator (Oracle). Results are based on 1000 simulated datasets, each with a sample size of 400.}
\centering
\resizebox{\linewidth}{!}{
\begin{tabular}[t]{llrrrrrrrrr}
\toprule
\multicolumn{2}{c}{\textbf{ }} & \multicolumn{3}{c}{\textbf{Percent Bias}} & \multicolumn{3}{c}{\textbf{Estimated SE}} & \multicolumn{3}{c}{\textbf{95\% Coverage}} \\
\cmidrule(l{5pt}r{5pt}){3-5} \cmidrule(l{5pt}r{5pt}){6-8} \cmidrule(l{5pt}r{5pt}){9-11}
\textbf{Censoring Rate} & \textbf{Method} & \textbf{$\wh{\beta}_0$} & \textbf{$\wh{\beta}_1$} & \textbf{$\wh{\beta}_2$} & \textbf{$\wh{\beta}_0$} & \textbf{$\wh{\beta}_1$} & \textbf{$\wh{\beta}_2$} & \textbf{$\wh{\beta}_0$} & \textbf{$\wh{\beta}_1$} & \textbf{$\wh{\beta}_2$}\\
\midrule
\addlinespace[0.3em]
\multicolumn{11}{c}{\textbf{Exogenous Censoring does not hold}}\\
 & CC & -13.01 & 0.43 & 0.29 & 0.16 & 0.09 & 0.08 & 0.94 & 0.94 & 0.94\\

\multirow[t]{-2}{*}{\raggedright\arraybackslash \hspace{1em}25\%} & Oracle & -0.78 & 0.24 & 0.16 & 0.14 & 0.08 & 0.07 & 0.95 & 0.95 & 0.94\\
\addlinespace
 & CC & -109.01 & 0.72 & 0.38 & 0.28 & 0.16 & 0.14 & 0.50 & 0.94 & 0.94\\

\multirow[t]{-2}{*}{\raggedright\arraybackslash \hspace{1em}75\%} & Oracle & -0.67 & 0.20 & 0.17 & 0.14 & 0.08 & 0.07 & 0.96 & 0.95 & 0.94\\
\addlinespace
\addlinespace[0.3em]
\multicolumn{11}{c}{\textbf{Strict Exogenous Censoring does not hold}}\\
 & CC & 0.35 & -0.07 & -0.11 & 0.19 & 0.11 & 0.09 & 0.94 & 0.94 & 0.94\\

\multirow[t]{-2}{*}{\raggedright\arraybackslash \hspace{1em}25\%} & Oracle & 0.16 & -0.04 & -0.08 & 0.14 & 0.08 & 0.07 & 0.94 & 0.95 & \vphantom{3} 0.93\\
\addlinespace
 & CC & -0.78 & -0.16 & -0.30 & 0.48 & 0.27 & 0.24 & 0.94 & 0.95 & 0.92\\

\multirow[t]{-2}{*}{\raggedright\arraybackslash \hspace{1em}75\%} & Oracle & 0.16 & -0.04 & -0.08 & 0.14 & 0.08 & 0.07 & 0.94 & 0.95 & \vphantom{3} 0.93\\
\addlinespace
\addlinespace[0.3em]
\multicolumn{11}{c}{\textbf{Conditional Independence given (X,Z) does not hold}}\\
 & CC & -1.38 & 0.44 & 0.18 & 0.15 & 0.09 & 0.08 & 0.96 & 0.96 & 0.94\\

\multirow[t]{-2}{*}{\raggedright\arraybackslash \hspace{1em}25\%} & Oracle & -0.70 & 0.19 & 0.11 & 0.14 & 0.08 & 0.07 & 0.95 & 0.94 & 0.94\\
\addlinespace
 & CC & -2.18 & 1.34 & 0.41 & 0.24 & 0.43 & 0.14 & 0.94 & 0.94 & 0.94\\

\multirow[t]{-2}{*}{\raggedright\arraybackslash \hspace{1em}75\%} & Oracle & -0.70 & 0.19 & 0.11 & 0.14 & 0.08 & 0.07 & 0.95 & 0.94 & 0.94\\
\addlinespace
\addlinespace[0.3em]
\multicolumn{11}{c}{\textbf{Conditional Independence given Z does not hold}}\\
 & CC & 0.11 & 0.14 & -0.10 & 0.16 & 0.09 & 0.08 & 0.94 & 0.94 & 0.93\\

\multirow[t]{-2}{*}{\raggedright\arraybackslash \hspace{1em}25\%} & Oracle & 0.16 & -0.04 & -0.08 & 0.14 & 0.08 & 0.07 & 0.94 & 0.95 & \vphantom{2} 0.93\\
\addlinespace
 & CC & -0.40 & 0.25 & -0.44 & 0.28 & 0.16 & 0.14 & 0.93 & 0.94 & 0.93\\

\multirow[t]{-2}{*}{\raggedright\arraybackslash \hspace{1em}75\%} & Oracle & 0.16 & -0.04 & -0.08 & 0.14 & 0.08 & 0.07 & 0.94 & 0.95 & \vphantom{2} 0.93\\
\addlinespace
\addlinespace[0.3em]
\multicolumn{11}{c}{\textbf{Independence does not hold}}\\
 & CC & 0.15 & 0.03 & -0.16 & 0.15 & 0.10 & 0.08 & 0.94 & 0.95 & 0.93\\

\multirow[t]{-2}{*}{\raggedright\arraybackslash \hspace{1em}25\%} & Oracle & 0.16 & -0.04 & -0.08 & 0.14 & 0.08 & 0.07 & 0.94 & 0.95 & \vphantom{1} 0.93\\
\addlinespace
 & CC & -0.50 & 0.70 & -0.45 & 0.23 & 0.24 & 0.15 & 0.93 & 0.92 & 0.92\\

\multirow[t]{-2}{*}{\raggedright\arraybackslash \hspace{1em}75\%} & Oracle & 0.16 & -0.04 & -0.08 & 0.14 & 0.08 & 0.07 & 0.94 & 0.95 & \vphantom{1} 0.93\\
\addlinespace
\addlinespace[0.3em]
\multicolumn{11}{c}{\textbf{Independence holds}}\\
 & CC & 0.14 & 0.00 & -0.18 & 0.15 & 0.10 & 0.08 & 0.94 & 0.95 & 0.93\\

\multirow[t]{-2}{*}{\raggedright\arraybackslash \hspace{1em}25\%} & Oracle & 0.16 & -0.04 & -0.08 & 0.14 & 0.08 & 0.07 & 0.94 & 0.95 & 0.93\\
\addlinespace
 & CC & 0.18 & 0.37 & -0.17 & 0.24 & 0.39 & 0.14 & 0.94 & 0.94 & 0.94\\

\multirow[t]{-2}{*}{\raggedright\arraybackslash \hspace{1em}75\%} & Oracle & 0.16 & -0.04 & -0.08 & 0.14 & 0.08 & 0.07 & 0.94 & 0.95 & 0.93\\
\bottomrule
\end{tabular}}
\end{table}

Whether the sample size was 400 (Table \ref{tab:sims_400_linear}) or 1200 (Table \ref{tab:sims_1200_linear}, Supplementary Materials \ref{app:sim_tabs}), we only see substantial empirical bias (i.e., bias $> 10\%$) when exogenous censoring does not hold, for a linear regression model. The bias under this setting congregates in the intercept estimate, $\wh{\beta}_0$, and gets worse with increasing censoring rates. The amount of bias in the intercept depends on how much the mean of the error term for $\Delta = 1$ differs from $0$ (See Supplementary Materials \ref{app:bias_linear_int}). Here, $E(\bepsilon|\bDelta = 1, \bX, \bZ) = -(\sigma/8)\times\{r/(1-r)\}$. We choose $\sigma/8$ arbitrarily; if the expectation were closer to $0$, there would be less bias, and if the difference were further than $0$, then the bias would be even worse. 

The bias for the parameter estimate associated with the censored covariate $\bX$, $\wh{\beta}_1$, however, is quite low across all settings (i.e., under $2\%$ bias). If the main goal of this analysis was to get an unbiased parameter estimate for $\beta_1$, then even a violation of exogenous censoring is okay, as long as the model has an intercept. The importance of the intercept is illustrated empirically in Table \ref{tab:sims_400_linear_int}, Supplementary Materials \ref{app:sim_tabs}, where we compare the percent bias for the $n=400$ simulation with the same simulation that has no intercept (i.e., the mean model is $m(X,Z;\bbeta) = X-2Z$). We see that when the intercept is removed, the bias transfers to the estimate for $\beta_1$ up to $27\%$, which is undesirable.

The precision of the estimates is a different story, as seen from the standard error estimates. Estimates from the complete case analysis increase in variability as the censoring rate increases and/or the sample size decreases due to a smaller effective sample size in both cases. For example, the complete case estimator leads to standard errors as high as 0.48 when $n = 400$, compared to 0.14 for the oracle estimator, and 0.25 when $n=1200$, compared to 0.08 for the oracle estimator. Similarly, when $n=400$, the largest standard error estimate when the censoring rate is $25\%$ is 0.19, as opposed to 0.48 for the $75\%$ censoring rate. The estimates also vary widely in precision across the different censoring mechanism settings, where we see standard error estimates for $\wh{\beta}_1$ range from 0.16 to 0.43 when $n=400$ and the censoring rate is $75\%$. The variation in precision is simply a consequence of how the data were generated in each setting. This result shows that despite the empirically unbiased complete case estimates, high censoring rates can still lead to imprecise estimates, depending on the underlying properties of the data. Despite the imprecision, however, we see that the coverage rates are close to the nominal $95\%$ for the complete case estimator in all settings where we have empirical unbiasedness. %We therefore expect the $95\%$ confidence interval for each estimate to contain the true value about $95\%$ of the time.

The results from the nonlinear regression model with a logistic mean function can be found in Supplementary Material \ref{app:sim_tabs} (Tables \ref{tab:sims_400_logistic} and \ref{tab:sims_1200_logistic}). We see the most substantial bias for $\wh{\beta}_0$ when exogenous censoring does not hold, ranging from $200\%$ to almost $2000\%$. For some censoring mechanism combinations, there is more bias from the $\wh{\beta}_1$ complete case estimator than from the oracle estimator. The bias could be due, in part, to the effective sample size of the complete case estimator being too small for consistency to ``kick in''. Illustrating this concept, we see bias as high as $24\%$ when $n=400$ and only $16\%$ when $n=1200$. In addition, the bias in $\wh{\beta}_1$ is likely due to the high variability we see in these estimates. For example, in the ``Conditional Independence given $Z$ does not hold'' setting, the standard error is seven times greater for complete case analysis than for oracle, when considering the $\wh{\beta}_1$ at $n=400$ and censoring rate $75\%$. We compare that relative efficiency to only three times greater in the linear case, where we see less bias.

%%%%%%%%%%%%%%%%%%%%%%%%%%%%
\section{Apathy in Huntington disease as it relates to clinical diagnosis time}
\label{sec:realdata}
%%%%%%%%%%%%%%%%%%%%%%%%%%%%%

\subsection{Motivation and Data}

Huntington disease is a neurodegenerative disease characterized by motor disturbances, cognitive decline, and psychiatric symptoms \cite{Langbehnetal2004}. Huntington disease is fully penetrant, meaning that if someone has the causal gene mutation, they are guaranteed to develop the disease in their lifetime. Individuals with the gene mutation and some symptoms, but not yet displaying enough motor disturbances to merit a clinical diagnosis, are in the ``prodromal'' disease stage \cite{Zhangetal2011}. Treating Huntington disease patients in the prodromal stage is important for possibly delaying diagnosis, avoiding irreparable damage and/or slowing symptom progression \cite{Paulsenetal2006}.

One of the most common prodromal symptoms is apathy, described as lacking feeling, emotion, interest, and/or concern \cite{Levy2006, Atkinsetal2021}. Apathy is distressing not only to the patient but also to the caretaker, since apathy presents as stubbornness and hostility which can strain relationships \cite{Kingmaetal2008, MasonBarker2015}. There is currently no treatment for apathy symptoms in Huntington disease, despite efforts to develop one \cite{Gelderblometal2017}. Therefore, one step in treating apathy as early as possible would be to characterize how it changes as a function of time to diagnosis, or disease progression, specifically \textit{within} the prodromal stage. Many researchers have concluded that apathy does indeed appear to be greater on average for those post-diagnosis than for those in the prodromal disease stage \cite{MartinezHortaetal2016, Fritzetal2018}. We are interested in determining how apathy changes over time to see where on the disease path apathy begins to increase or worsen. Knowing this information would enable researchers to target this time of initial apathy increase for future treatments and interventions.

The Neurobiological Predictors of Huntington Disease (PREDICT-HD) study is a longitudinal, observational study designed to explore symptoms of Huntington disease in the prodromal stage \cite{Paulsenetal2006}. In one study of apathy in the prodromal period using PREDICT-HD data \cite{Misiuraetal2019}, a cross-sectional linear regression analysis was performed with data from the baseline visit and apathy as the outcome. One variable \cite{Misiuraetal2019} included as a covariate in their model is a commonly used measure of disease burden called the CAG-Age-Product (CAP) score, which is a function of CAG repeat length and age \cite{Zhangetal2011}. The purpose of the CAP score is to predict ``time to clinical diagnosis'', or briefly, ``time to diagnosis''. Since \cite{Misiuraetal2019} did not see a significant relationship between CAP score and apathy while controlling for other variables, they concluded that apathy may not be \textit{directly} related to the progression of Huntington disease during the prodromal stage. This conclusion would mean that despite the literature arguing apathy increases across disease stages, the relationship may be more difficult to detect within the prodromal stage or may be confounded by other variables.

Given the longitudinal nature of PREDICT-HD, some of the participants were given a clinical diagnosis during the course of the study, and therefore their time to diagnosis was observed. Instead of using the CAP score to draw conclusions about the relationship between apathy and the progression of Huntington disease, we can use the patient's actual observed time to diagnosis. This approach is important because CAP scores are known to be imprecise and do not always accurately reflect clinical diagnosis \cite{Paulsenetal2019}. Therefore, running the analysis with the true time to diagnosis can more accurately assess whether the lack of relationship \cite{Misiuraetal2019} drew between apathy and disease progression is valid. The time to diagnosis covariate, however, is right-censored, since not everyone was given a clinical diagnosis by the time the study was over. This case presents an opportunity to analyze these data with a complete case estimator, which is consistent under one of the censoring mechanism assumptions \ref{assump:exogenous} \textendash \ref{assump:indep}. 

\subsection{Considering Censoring Mechanism Assumptions}

As discussed in Section \ref{sec:check-assump}, it is not possible to check whether censoring mechanism assumptions hold based on the observed data alone, due to the censoring of $\bX$. That being said, the dropout rate for PREDICT-HD was less than $5\%$ per year, and sample size variation year to year was due to study design considerations rather than dropouts \cite{Paulsenetal2014a}. Keeping this in mind, we can look back at the DAG in Figure \ref{fig:cc_dag_a} to revisit our proposed relationships between these variables. We remember from Section \ref{sec:check-assump} that $\bY \independent \bC | (\bX, \bZ)$ as long as there is no direct path between $\bY$ and $\bC$. We mentioned the possibility that apathetic people may lose interest in the study and dropout, causing a relationship between apathy and the time to censoring. However, given such a low dropout rate, we are confident in assuming this path does not exist in the true DAG. Therefore, we assume censoring mechanism \ref{assump:condxz}, which will lead to a consistent complete case estimator.

\subsection{Analysis}

We focus on analyzing individuals who were genetically confirmed to have the gene mutation for Huntington disease (i.e., a CAG repeat length of at least 36). We also remove anyone who was given a clinical diagnosis at the first visit because we wanted a subset of individuals in the prodromal stage at baseline. Furthermore, we run the analysis using data from the second visit, since some patients were diagnosed at the second visit, giving us a wider range for the time to diagnosis variable. Finally, we remove any individuals with \textit{missing} outcome or covariate information, leaving us with an analysis set of 781, where 212 have an observed time to diagnosis. This is a censoring rate of about $73\%$. 

The goal of this analysis is to quantify the linear association between apathy and time to clinical diagnosis (in years), while controlling for sex, depression, total motor score, years of education, single digit modalities test score (SDMT), CAG repeat length, and age, similar to the model construction by \cite{Misiuraetal2019}. Apathy was measured using the Frontal Systems Behavior Scale, which measures behavior associated with frontal systems of the brain and has subscales for apathy, disinhibition, and executive function \cite{Grace2018}. Depression was measured using the Symptom Checklist-90-R, which measures psychological distress and has several subscales including depression \cite{DerogatisUnger2010}. Both the patient and caretaker filled out these rating scales and we use the caretaker results. We chose a linear model to reflect the relationship between apathy and disease progression from previous studies \cite{MartinezHortaetal2016, Misiuraetal2019}. This set of covariates has been shown to be related to apathy, and we want to control for them to see if time to diagnosis is directly related to apathy, rather than through a relationship with these control variables. We want to compare our results to the model from \cite{Misiuraetal2019} to see if using time to diagnosis yields different results than the CAP score. If the results agree, we will conclude that there is no evidence for a strong, direct relationship between apathy and disease progression during the prodromal stage. If the results are different, there will be some evidence to suggest that apathy does change with disease progression when the true time to diagnosis variable is used.

We use the following linear regression model to test these relationships: 
\begin{align*}
  Y_i = \beta_0 + \beta_1X_i + \beta_2Z_{1i} + \beta_3Z_{2i} + \beta_4Z_{3i} + \beta_5Z_{4i} + \beta_6Z_{5i} + \beta_7Z_{6i} + \beta_8Z_{7i} + \epsilon_i, 
\end{align*}
where $Y_i$ is apathy score, $X_i$ is time to diagnosis in years, $Z_{1i}$ is biological sex indicator (1 if female), $Z_{2i}$ is depression score, $Z_{3i}$ is total motor score, $Z_{4i}$ is number years of education, $Z_{5i}$ is SDMT score, $Z_{6i}$ is CAG repeat length and $Z_{7i}$ is age in years, all at the second visit. All covariates, except time to diagnosis, were centered and scaled.

The results from the regression (Table \ref{tab:hd_analysis_ttd}) show no evidence that our measure of disease progression (i.e., time to diagnosis) is associated with apathy after controlling for other variables. These results agree with the same regression replacing time to diagnosis with CAP score (Table \ref{tab:hd_analysis_cap}, Supplementary Materials \ref{app:sim_tabs}) and the results from the CAP score model in \cite{Misiuraetal2019}. Therefore, our results agree with the conclusion made from the CAP score model, even if using CAP score may not have been a precise measure of disease progression. It is also worth noting that the only covariate whose confidence interval does not contain the null value in our time to diagnosis model is depression. The \cite{Misiuraetal2019} model and our CAP score model also saw evidence of a relationship between apathy and total motor score, as well as cognitive control (SDMT score, here). One reason for this result could be the lack of efficiency that comes with a complete case analysis. Because we deleted so much data for the time to diagnosis analysis, our standard errors are quite large, despite being confident in having a consistent estimator. Overall, our analysis shows that it may be difficult to detect a change in apathy scores in Huntington disease patients during the prodromal change, despite the clear increase across disease stage groups presented in the literature. From these results, we cannot recommend a clear way to recruit patients for trials that would be expected to detect a change in apathy over time. 

\begin{table}[htbp]

\caption{\label{tab:hd_analysis_ttd} Results from the PREDICT-HD study quantifying the relationship between apathy score and time to clinical diagnosis, controlling for other covariates. The parameters are estimated using a complete case analysis (n = 212). We report the parameter estimate, the estimated standard error (SE), and the Wald-type 95\% confidence interval (95\% CI).
}
\centering
\begin{tabular}[t]{rrrr}
\toprule
\textbf{Parameter} & \textbf{Estimate} & \textbf{Estimated SE} & \textbf{$95\%$ CI} \\
\midrule
\addlinespace[0.3em]

$\beta_0:$ Intercept & 15.02 & 0.96 & (13.15, 16.90) \\
$\beta_1:$ Time to Diagnosis & -0.015 & 0.15 & (-0.30, 0.27) \\
$\beta_2:$ Biological Sex & -1.38 & 0.86 & (-3.05, 0.30) \\
$\beta_3:$ Depression & 4.16 & 0.38 & (3.41, 4.91) \\
$\beta_4:$ Total Motor Score& 0.61 & 0.38 & (-0.13, 1.36) \\
$\beta_5:$ Years of Education & 0.30 & 0.46 & (-0.60, 1.21) \\
$\beta_6:$ SDMT Score & -0.24 & 0.50 & (-1.23, 0.75) \\
$\beta_7:$ CAG Repeat Length & 0.58 & 0.54 & (-0.48, 1.65) \\
$\beta_8:$ Age & -0.36 & 0.51 & (-1.36 , 0.64) \\

\bottomrule

\end{tabular}
\end{table}

\section{Discussion}
\label{sec:discussion}
%%%%%%%%%%%%%%%%%%%%%%%%%%%%%%%

This paper provides a unified summary of complete case estimator consistency for censored covariates, and we recommend that these properties are taken into consideration when using the method in any scenario. Despite its main drawback of lack of precision due to the discarding of data, the complete case estimator can serve as a powerful initial estimator for complex methods or as a baseline comparison when benchmarking novel methods. Additionally, using the complete case estimator does not require any model assumptions on the censored covariate, meaning there is less chance of model misspecification leading to bias. 

We show here that for regression models with a right-censored covariate, the complete case estimator is consistent when any one of the five censoring mechanisms \ref{assump:exogenous} through \ref{assump:indep} hold. These assumptions revolve around the idea that the censoring variable (in terms of $\bC$ or $\bDelta$) is independent of the outcome $\bY$, conditional or unconditional on covariates $\bX$ and $\bZ$. Moreover, a consistent estimator is not guaranteed when the exogenous censoring assumption does not hold. Although it is impossible to verify the assumptions made, domain knowledge and DAGs can be used to help gain confidence. Lastly, while we presented results when a covariate is randomly right-censored, our conclusions similarly apply when the covariate is left-censored (i.e., $\Delta = I(X \geq C$)) or when the censoring is due to a limit of detection (i.e., mechanisms \ref{assump:exogenous} \textendash \ref{assump:strictexogenous} only because there is no random $C$).

%\bigskip
\begin{center}
{\large\bf SUPPLEMENTARY MATERIAL}
\end{center}
The supplementary material contains all proofs, additional simulation tables and results for the Huntington disease example. R code for the simulation study is provided at \href{https://github.com/marissaashner/Complete_Case_Censored_Covariates}{https://github.com/marissaashner/Complete\_Case\_Censored\_Covariates}.

\section*{Acknowledgments}
This material is based upon work supported by the National Science Foundation Graduate Research Fellowship under grant DGE-2040435, the National Institute of Environmental Health Sciences under grant P30ES010126, and the National Institute of Neurological Disorders and Stroke under grant K01NS099343. The authors thank PREDICT-HD for permission to present their data.

%Bibliography
\bibliographystyle{unsrt}  
\bibliography{references}  

\newpage

  \begin{center}
  {\LARGE\bf Supplementary Material to \textit{Exploring the validity of the complete case analysis for regression models with a right-censored covariate}}
\end{center}

\newpage
%{\appendix\def\thesection{Appendix~S.\arabic{section}}\def\thesubsection{\Alph{section}.\arabic{subsection}}}
%\renewcommand{\theequation}{S.\arabic{equation}}
\renewcommand{\thesection}{S.\arabic{section}}
\setcounter{section}{0}
\setcounter{equation}{0}

\section{Proof of Theorem 1}
\label{app:th1}
We prove Theorem \ref{thm:censoring-relationships} by proving four implications between the five censoring mechanisms. Each of these proofs is labeled below, where `$\implies$' means `implies'.

\textbf{Censoring mechanism \ref{assump:indep} $\implies$ Censoring mechanism \ref{assump:condz}:} 
Assuming censoring mechanism \ref{assump:indep} holds, i.e., $\bC$ is independent of $(\bX, \bY, \bZ)$, we have that 
\bse
f_{\bC, \bZ}(c, \bz) &=& \int_{-\infty}^{\infty} \int_{-\infty}^{\infty} f_{\bC,\bY,\bX,\bZ}(c, y, x,\bz)dydx = \int_{-\infty}^{\infty} \int_{-\infty}^{\infty} f_{\bC}(c) f_{\bY, \bX, \bZ}(y,x,\bz)dydx
\\ 
&=& 
f_{\bC}(c) \int_{-\infty}^{\infty} \int_{-\infty}^{\infty} f_{\bY, \bX, \bZ}(y,x,z) dydx= f_{\bC}(c) f_{\bZ}(z). \ese

The result above shows that the joint density of $\bC$ and $\bZ$ equals the product of their marginal distributions, which means $\bC$ and $\bZ$ are marginally independent. This marginal independence combined with censoring mechanism  \ref{assump:indep} further implies that
\bse
f_{(\bC, \bY, \bX)|\bZ}(c,y,x|\bz) &=& \frac{f_{\bC, \bY, \bX, \bZ}(c,y,x,\bz)}{f_{\bZ}(\bz)}\\
&=&\frac{f_{\bC}(c)f_{\bY, \bX, \bZ}(y,x,\bz)}{f_{\bZ}(\bz)}\\
&=& f_{\bC}(c)f_{\bY, \bX| \bZ}(y,x|\bz)\\
&=& f_{\bC|\bZ}(c|\bz)f_{\bY, \bX| \bZ}(y,x|\bz).
\ese

The second equality above holds because we assumed censoring mechanism \ref{assump:indep}, %$\bC$ is assumed to be independent of $(\bY,\bX,\bZ)$, 
and the last equality holds because $\bC$ and $\bZ$ are marginally independent. We therefore showed $f_{(\bC, \bY, \bX)|\bZ}(c,y,x|z)=f_{\bC|\bZ}(c|\bz)f_{\bY, \bX| \bZ}(y,x|z)$, which proves $\bC$ is independent of $(\bX, \bY)$ given $\bZ$.

\textbf{Censoring mechanism \ref{assump:condz} $\implies$ Censoring mechanism  \ref{assump:condxz}:} 
Assuming censoring mechanism \ref{assump:condz}, i.e., $\bC$ is independent of $(\bX, \bY)$ given $\bZ$, we have that \bse
f_{\bC, \bX| \bZ}(c,x|\bz) &=& \int_{-\infty}^{\infty}  f_{\bC,\bY,\bX|\bZ}(c,y,x|\bz)dy = \int_{-\infty}^{\infty} f_{\bC|\bZ}(c|\bz) f_{\bY, \bX| \bZ}(y,x|\bz)dy 
\\
&=&
f_{\bC|\bZ}(c|\bz) \int_{-\infty}^{\infty} f_{\bY, \bX| \bZ}(y,x|\bz) dy = f_{\bC|\bZ}(c|\bz) f_{\bX| \bZ}(x|\bz).
\ese
Having shown $f_{\bC, \bX| \bZ}(c,x|\bz)=f_{\bC|\bZ}(c|\bz) f_{\bX| \bZ}(x|\bz)$, we proved that $\bC$ and $\bX$ are  independent conditional on $\bZ$. This conditional independence combined with censoring mechanism  \ref{assump:condz} implies that 
\bse
f_{\bC, \bY|\bX, \bZ}(c,y|\bx,\bz) &=& \frac{f_{\bC, \bY, \bX| \bZ}(c,y,x|\bz)}{f_{\bX| \bZ}(x|\bz)}
\\
&=&
\frac{f_{\bC|\bZ}(c|\bz)f_{\bY, \bX|\bZ}(x,y|z)}{f_{\bX|\bZ}(x|\bz)}
\\
&=&
\frac{f_{\bC|\bZ}(c|\bz)f_{\bY|\bX, \bZ}(y|x,\bz)f_{\bX| \bZ}(x|\bz)}{f_{\bX| \bZ}(x|\bz)}
\\
&=&
f_{\bC|\bZ}(c|\bz)f_{\bY|\bX,\bZ}(y|x,\bz)
\\
&=&
f_{\bC|\bX, \bZ}(c|x,\bz)f_{\bY|\bX, \bZ}(y|x,\bz).
\ese

The second equality holds by censoring mechanism \ref{assump:condz}, and the last equality holds because $\bC$ and $\bX$ are independent given $\bZ$. Because we showed that the conditional density of $(\bC,\bY)|\bX,\bZ$ is equal to the product of the conditional densities of $\bC|\bX,\bZ$ and $\bY|\bX,\bZ$, we proved $\bC$ is independent of $\bY$ given $(\bX, \bZ)$. 

\textbf{Censoring mechanism \ref{assump:condxz} $\implies$ Censoring mechanism \ref{assump:strictexogenous}:}
Let $F_V(v) = P(V\leq v)$ denote the cumulative distribution function for a random variable $V$. Assuming censoring mechanism \ref{assump:condxz}, i.e., $\bC$ is independent of $\bY$ given $(\bX, \bZ)$, we have that 
\bse
F_{\bC, \bY|\bX, \bZ}(c, y|X = x, \bZ = \bz) = F_{\bC|\bX,\bZ}(c|X=x, \bZ=\bz) F_{\bY|\bX,\bZ}(y|X=x, \bZ=\bz),
\ese
for all $c, y, x, \bz$. If we show that 
\be
\label{eqn:main-claim}
F_{\bDelta, \bepsilon|\bX, \bZ}(d, e|X = x, \bZ = \bz) = F_{\bDelta|\bX,\bZ}(d|X=x, \bZ=\bz) F_{\bepsilon|\bX,\bZ}(e|X=x, \bZ=\bz) 
\ee
for all $d, e, x, \bz$, we will have proven that censoring  mechanism \ref{assump:strictexogenous} holds. We will prove equation \eqref{eqn:main-claim} holds by considering three cases for $d\in \mathbb{R}$:
\begin{enumerate}
    \item Case 1: $d<0$. Because $\bDelta$  can only be $0$ or $1$ with nonzero probability, then it immediately follows that $F_{\bDelta, \bepsilon|\bX, \bZ}(d, e|X = x, \bZ = \bz)=0$ and $ F_{\bDelta|\bX,\bZ}(d|X = x, \bZ=\bz)=0$. Therefore, equation \eqref{eqn:main-claim} holds trivially.
    
    \item {Case 2: $0 \leq d < 1$.} When $0 \leq d < 1$, the only plausible value for $\bDelta$  that gives a non-zero probability  is $d=0$. Therefore, 
    \bse
    &&F_{\bDelta, \bepsilon|\bX, \bZ}(d, e|X = x, \bZ = \bz)\\
    &=& P(\bDelta \leq d, \bepsilon \leq e |X = x, \bZ = \bz)
    \\
    &=&
    P\{\bDelta = 0, \bY - m(\bX,\bZ;\bbeta) \leq e |X = x, \bZ = \bz\} 
    \\
    &=&
    P(\bC \leq \bX| X = x, \bZ = \bz)P\{\bY \leq m(\bX,\bZ;\bbeta) + e |X = x, \bZ = \bz\}
    \\
    &=&
    P(\bDelta = 0| X = x, \bZ = \bz)P(\bepsilon \leq e |X = x, \bZ = \bz) 
    \\
    &=&
    P(\bDelta\leq d |X = x, \bZ = \bz)P(\bepsilon \leq e |X = x, \bZ = \bz)
    \\
    &=&
    F_{\bDelta|\bX,\bZ}(d|X = x, \bZ=\bz) F_{\bepsilon|\bX,\bZ}(e|X = x, \bZ=\bz),
    \ese

where the third equality holds by censoring mechanism \ref{assump:condxz}. 
    \item Case 3: $d \geq 1$. 
    When $d \geq 1$, the plausible values for $\bDelta$  that gives a non-zero probability are $d=0$ and $d=1$. Therefore, 
    \bse
    F_{\bDelta, \bepsilon|\bX, \bZ}(d, e|X = x, \bZ = \bz) &=& P(\bDelta \leq d, \bepsilon \leq e |X = x, \bZ = \bz)
    \\
    &=&
    P(\bDelta =0 \text{ or } 1, \bepsilon \leq e |X = x, \bZ = \bz) 
    \\
    &=&
    1\times P(\bepsilon \leq e |X = x, \bZ = \bz)
    \\
    &=&
    P(\bDelta\leq d |X = x, \bZ = \bz)P(\bepsilon \leq e |X = x, \bZ = \bz)
    \\
    &=&
    F_{\bDelta|\bX,\bZ}(d|X = x, \bZ=\bz) F_{\bepsilon|\bX,\bZ}(e|X = x, \bZ=\bz).
    \ese    
\end{enumerate}
%We have shown that $F_{\bDelta, \bepsilon|\bX, \bZ}(d, e|X = x, \bZ = z) = F_{\bDelta\bX,\bZ}(d|X = x, \bZ=z) F_{\bepsilon|\bX,\bZ}(e|X = x, \bZ=z)$ for all $d \in \mathbb{R}$, and for all $e, x,z \in \mathbb{R}$, since these were unspecified in each case. Therefore, $\bDelta$ is independent of $\bepsilon$ given $(\bX, \bZ)$. 

\textbf{Censoring mechanism \ref{assump:strictexogenous} $\implies$ Censoring mechanism \ref{assump:exogenous}:} Assuming censoring mechanism \ref{assump:strictexogenous}, i.e., $\bDelta$ is independent of $\bepsilon$ given $(\bX, \bZ)$, we have
\bse
E(\bepsilon|\bDelta, \bX, \bZ) = E(\bepsilon|\bX,\bZ).
\ese 
But because we assume $E(\bepsilon|\bX,\bZ)=\bzero$, we have that censoring mechanism \ref{assump:exogenous} holds.

\section{Proof of Theorem 2}
\label{app:consistent_cc_nonlinear}

\textbf{Consistency: } Following the work of  \cite{YuanJennrich1998} and \cite{Tsiatis2006}, consistency will follow so long as we show that the following three regularity conditions hold, under the assumption that \ref{reg:unbias}\textendash\ref{reg:uniform} hold for the fully observed estimating equation approach, as laid out in Theorem \ref{thm:exogenous-consistent} in the main text.
\begin{enumerate}[label = (R\arabic*$^*$)]
    \item \label{reg:unbias_cc}  $E\{\bDelta\bPhi(\bY, X, \bZ;\bbeta)\} = \bzero$ for all $\bbeta$
    \item \label{reg:nonsing_cc} $E\left\{\frac{\partial\bDelta\bPhi(\bY, \bX, \bZ;\bbeta)}{\partial \bbeta^T}\right\}$ is nonsingular 
    \item \label{reg:uniform_cc} $n^{-1} \sum_{i=1}^n \frac{\partial\Delta_i\bPhi(Y_i, X_i, \bZ_i;\bbeta)}{\partial \bbeta^T}$ converges uniformly to $E\{\frac{\partial\bDelta\bPhi(\bY, X, \bZ;\bbeta)}{\partial \bbeta^T}\}$ in a neighborhood of $\bbeta_0$, where $\bbeta_0$ is the true parameter value.
\end{enumerate}

 \textbf{\ref{reg:unbias_cc}:} First, we show $E\{\bDelta\bPhi(\bY, \bX, \bZ;\bbeta)\} = \bzero$ for all $\bbeta$. Observe that 
\bse
&&E\{\bDelta\bPhi(Y,\bX,\bZ;\bbeta)\} = E[\Delta_i\bA(X,\bZ;\bbeta)\{Y-m(X,\bZ;\bbeta)\}]
\\
&=&
  E_{\bX,\bZ, \bDelta}\{E_{Y|\bX,\bZ, \bDelta}[\Delta_i\bA(X,\bZ;\bbeta)\{Y-m(X,\bZ;\bbeta)\}|\bX,\bZ, \bDelta]\}
\\
&=&
    E_{\bX,\bZ, \bDelta}[\Delta_i\bA(X,\bZ;\bbeta)\{E_{Y|\bX,\bZ, \bDelta}(Y|\bX,\bZ, \bDelta)-m(X,\bZ;\bbeta)\}]
\\
&=&
E_{\bX,\bZ, \bDelta}(\Delta_i\bA(X,\bZ;\bbeta)[E_{Y|\bX,\bZ, \bDelta}\{\epsilon + m(X,\bZ;\bbeta)|\bX,\bZ, \bDelta\}-m(X,\bZ;\bbeta)])
\\ 
&=& 
E_{\bX,\bZ, \bDelta}(\Delta_i\bA(X,\bZ;\bbeta)[E_{Y|\bX,\bZ, \bDelta}\{\epsilon|\bX,\bZ, \bDelta\} + m(X,\bZ;\bbeta)-m(X,\bZ;\bbeta)]), \ese
where the second equality holds by iterated expectations conditioning on $\bDelta, \bX, \bZ$.
Since we are conditioning on $\bDelta, \bX, \bZ$, we can pull any terms dependent on these variables out of the inner expectation in the third equality. From the last equality, the $m(X,\bZ;\bbeta)$ will cancel out, leaving the inner expectation which equals $\bzero$ by exogenous censoring. Therefore, $E\{\bDelta\bPhi(\bY, \bX, \bZ;\bbeta)\} = \bzero$. 

 \textbf{\ref{reg:nonsing_cc}:}  Now, we will show $E\left\{\frac{\partial\bDelta\bPhi(\bY, \bX, \bZ;\bbeta)}{\partial \bbeta^T}\right\}$ is nonsingular. By regularity condition \ref{reg:nonsing}, we have that  $E\left\{\frac{\partial\bPhi(\bY, \bX, \bZ;\bbeta)}{\partial \bbeta^T}\right\}$ is nonsingular. Then,
\bse
E\left\{\frac{\partial\bPhi(\bY, \bX, \bZ;\bbeta)}{\partial \bbeta^T}\right\} &=& E\left[\frac{\partial \bA(\bX, \bZ;\bbeta)\{\bY-m(\bX,\bZ;\bbeta)\}}{\partial \bbeta^T}\right]
\\
&=&
E\left[\frac{\partial\bA(\bX,\bZ;\bbeta)}{\partial\bbeta}\{\bY-m(\bX,\bZ;\bbeta)\} - \bA(\cdot)\frac{\partial m(\bX,\bZ;\bbeta)}{\partial\bbeta}\right],
\ese
where the second equality holds by the product rule for derivatives.
Consider the first term, $E\left[\partial\bA(\bX,\bZ;\bbeta)/\partial\bbeta\{\bY-m(\bX,\bZ;\bbeta)\}\right]$. Using iterated expectations, this is equal to 
\bse
 &=& E_{\bX,\bZ}\left(E_{\bY|\bX,\bZ}\left[\frac{\partial\bA(\bX,\bZ;\bbeta)}{\partial\bbeta}\{\bY-m(\bX,\bZ;\bbeta)\}|\bX,\bZ\right]\right)
 \\
 &=&
 E_{\bX,\bZ}\left\{\frac{\partial\bA(\bX,\bZ;\bbeta)}{\partial\bbeta}E_{\bY|\bX,\bZ}\left(\bepsilon|\bX,\bZ\right)\right\},
\ese
and the inner expectation is $\bzero$ by the original model assumption on the error term. This means our assumption on the full data case simplifies to nonsingularlity of the following matrix: $-E\left\{ \bA(\bX,\bZ;\bbeta){\partial m(\bX,\bZ;\bbeta)}/{\partial\bbeta}\right\}.$ 

We can simplify the expectation of the complete case estimating function the same way, and the first term will also equal $\bzero$ via iterative expectations and exogenous censoring. Then we want to show the nonsingularity of $-E\left\{ \bDelta\bA(\bX,\bZ;\bbeta){\partial m(\bX,\bZ;\bbeta)}/{\partial\bbeta}\right\}.$ Observe that this expectation is equal to 
\bse
 &=& -\int_{-\infty}^{\infty}\int_{-\infty}^{\infty}\int_{-\infty}^{\infty} \delta \bA(x, z;\bbeta) \frac{\partial m(x,z;\bbeta)}{\partial\bbeta} f_{\bX,\bZ,\bDelta}(x,z,\delta) d\delta dx dz 
\\
&=&
-\int_{-\infty}^{\infty}\int_{-\infty}^{\infty}  \bA(x, z;\bbeta) \frac{\partial m(x,z;\bbeta)}{\partial\bbeta} f_{\bX,\bZ}(x,z) \left\{\int_{-\infty}^{\infty}\delta f_{\bDelta|\bX,\bZ}(\delta|x,z) d\delta\right\} dx dz
\\
&=&
E\left\{P(\bDelta=1|\bX,\bZ)\bA(\bX,\bZ;\bbeta)\frac{\partial m(\bX,\bZ;\bbeta)}{\partial\bbeta}\right\},
\ese
where the second equality holds since $f_{\bX,\bZ,\bDelta}(x,z,\delta) = f_{\bDelta|\bX,\bZ}(\delta|x,z)f_{\bX,\bZ}(x,z)$, and the third equality holds since $\int_{-\infty}^{\infty}\delta f_{\bDelta|\bX,\bZ}(\delta|x,z) d\delta = P(\bDelta=1|\bX,\bZ)$. This can be rewritten as $E\left\{\bA^*(\bX,\bZ;\bbeta){\partial m(\bX,\bZ;\bbeta)}/{\partial\bbeta}\right\}$, where $A^*(\bX,\bZ;\bbeta) = P(\bDelta=1|\bX,\bZ)\bA(\bX,\bZ;\bbeta)$. Since we assumed the original matrix was nonsingular for arbitrary $\bA(\bX,\bZ;\bbeta)$, it is nonsingular for $A^*(\bX,\bZ;\bbeta)$.

 \textbf{\ref{reg:uniform_cc}:}  Finally, we will show $n^{-1} \sum_{i=1}^n {\partial\Delta_i\bPhi(Y_i, X_i, \bZ_i;\bbeta)}/{\partial \bbeta^T}$ converge uniformly to $E\left\{\frac{\partial\bDelta\bPhi(\bY, \bX, \bZ;\bbeta)}{\partial \bbeta^T}\right\}$ in a neighborhood of $\bbeta_0$, where $\bbeta_0$ is the true parameter value. As in \cite{Tsiatis2006}, uniform convergence would be satisfied if sample paths of $\partial\Delta\bPhi(Y, X, \bZ;\bbeta)/\partial \bbeta^T$ are continuous in $\bbeta$ about $\bbeta_{0}$ almost surely and $$\sup_{\bbeta \in \mathcal{N}(\bbeta_{0})} \left | \frac{\partial \Delta\bPhi(Y, X, \bZ;\bbeta)}{\partial\bbeta^T}\right | \leq g(\bY, \bX, \bZ, \bC),$$ 
where $E\{g(\bY, \bX, \bZ, \bC)\} < \infty$ and $\mathcal{N}(\bbeta_{0})$ denotes a neighborhood in $\bbeta$ about $\bbeta_{0}$.

Remember our assumption that uniform convergence holds for the full data setting. More specifically, using this last definition, this means that sample paths of $\partial \bPhi/\partial \bbeta^T$ are continuous in $\bbeta$ about $\bbeta_{0}$ almost surely and $$\sup_{\bbeta \in \mathcal{N}(\bbeta_{0})} \left | \frac{\partial \bPhi(\bY,\bX,\bZ;\bbeta)}{\partial\bbeta^T}\right | \leq g^*(\bY, \bX, \bZ),$$ 
where $E\{g^*(\bY, \bX, \bZ)\} < \infty$ and $\mathcal{N}(\bbeta_{0})$ denotes a neighborhood in $\bbeta$ about $\bbeta_{0}$. We now want to show that uniform convergence holds when we replace $\bPhi$ with $\bDelta\bPhi$. Consider
\bse
    g^*(\bY, \bX, \bZ) &\geq& 
    \sup_{\bbeta \in \mathcal{N}(\bbeta_{0})} \left | \frac{\partial \bPhi(\bY, \bX, \bZ;\bbeta)}{\partial\bbeta^T}\right | 
    \\
    &\geq&
    \sup_{\bbeta \in \mathcal{N}(\bbeta_{0})} \left | \frac{\Delta \partial \bPhi(\bY, \bX, \bZ;\bbeta)}{\partial\bbeta^T}\right |,
\ese
where the second inequality holds since $\bDelta$ can only take on values $0$ or $1$. Since $\bDelta$ is a parameter that is not a function of $\bbeta$, we can move it inside the derivative as such: 
\bse
\sup_{\bbeta \in \mathcal{N}(\bbeta_{0})} \left | \frac{\Delta\partial \bPhi(\bY, \bX, \bZ;\bbeta)}{\partial\bbeta^T}\right | = \sup_{\bbeta \in \mathcal{N}(\bbeta_{0})} \left | \frac{\partial \Delta\bPhi(\bY, \bX, \bZ;\bbeta)}{\partial\bbeta^T}\right |.
\ese
Let $g(\bY, \bX, \bZ, \bC)= g^*(\bY, \bX, \bZ)$. Then we've shown $$\sup_{\bbeta \in \mathcal{N}(\bbeta_{0})} \left | \frac{\partial \Delta\bPhi(Y, X, \bZ;\bbeta)}{\partial\bbeta^T}\right | \leq g(\bY, \bX, \bZ, \bC).$$

\textbf{Asymptotic Normality: } Now, we want to show that the asymptotic distribution of $\wh{\bbeta}$ is $$\sqrt{n}(\wh{\bbeta}-\bbeta_0) \rightarrow \Normal(\bzero, \bA^{-1}\bB(\bA^{-1})^T),$$
where $\bA = E\{\partial \Delta\bPhi(Y,X,\bZ;\bbeta_0)/\partial\bbeta^T\}$ and $\bB = E[\{\Delta\bPhi(Y,X,\bZ;\bbeta_0)\}\{\Delta\bPhi(Y,X,\bZ;\bbeta_0)\}^T]$. This proof will follow similarly to that of \cite{Tsiatis2006}. First, we take the first order Taylor expansion of the estimating equation about $\bbeta_0$:
\bse
\bzero = \sum_{i=1}^n \Delta_i\bPhi(Y_i,X_i,\bZ_i;\wh{\bbeta}) = \sum_{i=1}^n \Delta_i\bPhi(Y_i,X_i,\bZ_i;\bbeta_0) + \left\{\sum_{i=1}^n\frac{\partial \Delta_i\bPhi(Y_i,X_i,\bZ_i;\bbeta^*)}{\partial\bbeta^T}\right\}(\wh{\bbeta}-\bbeta_0),
\ese
where $\bbeta^*$ is an intermediate value between $\wh{\bbeta}$ and $\bbeta_0$. Given our nonsingularity assumption, we then isolate the difference between $\wh{\bbeta}$ and $\bbeta_0$ such that 
\bse
\sqrt{n}(\wh{\bbeta}-\bbeta_0) = -\left\{\frac{1}{n}\sum_{i=1}^n\frac{\partial \Delta_i\bPhi(Y_i,X_i,\bZ_i;\bbeta^*)}{\partial\bbeta^T}\right\}^{-1}\frac{1}{\sqrt{n}}\sum_{i=1}^n \Delta_i\bPhi(Y_i,X_i,\bZ_i;\bbeta_0).
\ese

Now, note that, by the law of large numbers, 
\bse
 -\left\{\frac{1}{n}\sum_{i=1}^n\frac{\partial \Delta_i\bPhi(Y_i,X_i,\bZ_i;\bbeta^*)}{\partial\bbeta^T}\right\}^{-1} \rightarrow_p -\left[E\left\{\frac{\partial\Delta_i\bPhi(Y_i,X_i,\bZ_i;\bbeta_0}{\partial\bbeta}\right\}\right],
\ese
and by the Central Limit Theorem,
\bse
\frac{1}{\sqrt{n}}\sum_{i=1}^n \Delta_i\bPhi(Y_i,X_i,\bZ_i;\bbeta_0) &=& \sqrt{n}\times (\frac{1}{n}\sum_{i=1}^n \Delta_i\bPhi(Y_i,X_i,\bZ_i;\bbeta_0)-\bzero) 
\\
&\rightarrow_d& \Normal(\bzero,Var\{\Delta\bPhi(Y,X,\bZ;\bbeta_0)\}),
\ese

where $Var\{\Delta\bPhi(Y,X,\bZ;\bbeta_0)\} = E[\{\Delta\bPhi(Y,X,\bZ;\bbeta_0)\}\{\Delta\bPhi(Y,X,\bZ;\bbeta_0)\}^T]$. Finally, by Slutsky's Theorem,
\bse
\sqrt{n}(\wh{\bbeta}-\bbeta_0) \rightarrow_d \Normal(\bzero, \bA^{-1}\bB(\bA^{-1})^T),
\ese
where $\bA = E\{\partial \Delta\bPhi(Y,X,\bZ;\bbeta_0)/\partial\bbeta^T\}$ and $\bB = E[\{\Delta\bPhi(Y,X,\bZ;\bbeta_0)\}\{\Delta\bPhi(Y,X,\bZ;\bbeta_0)\}^T]$.

\section{Investigation of Bias for Linear Model}
\label{app:bias_linear_int}

In this section, we explore the bias that can occur for a complete case analysis with a linear model if exogenous censoring \ref{assump:exogenous} does not hold. This follows the discussion from Section \ref{sec:bias_linear}. Without loss of generality, we will show the simplest case where there is an intercept but no other $\bZ$ covariates such that $m(\cdot) = \beta_0 + \beta_1 X$. The estimating equation from Section \ref{sec:bias_linear} will have two components: 
\begin{enumerate}
    \item $\frac{1}{n}\sum_{i=1}^n \Delta_i(Y_i-\beta_0-\beta_1X_i) = 0$ and 
    \item $\frac{1}{n}\sum_{i=1}^n \Delta_iX_i(Y_i-\beta_0-\beta_1X_i) = 0$.
\end{enumerate}
Considering the first component, we can isolate $\beta_0$ to get $\wh{\beta_0} = \sum_{i=1}^n \Delta_i(Y_i-\wh{\beta}_1X_i) / \sum_{i=1}^n \Delta_i$. Then, plugging $\wh{\beta}_0$ in for $\beta_0$ in the second component, we can isolate $\beta_1$ to get $\wh{\beta}_1 = \sum_{i=1}^n \Delta_iX_i(Y_i-\sum_{i=1}^n \Delta_iY_i/\sum_{i=1}^n \Delta_i) / \{\sum_{i=1}^n \Delta_iX_i^2 - (\sum_{i=1}^n  \Delta_iX_i)^2/\sum_{i=1}^n\Delta_i\}$.

We will show that $\wh{\beta}_1$ is unbiased, even if exogenous censoring is broken, by showing the expectation is equal to $\beta_1$. 
\bse
E(\wh{\beta}_1) &=& E\left\{\frac{\sum_{i=1}^n \Delta_iX_i(Y_i-\sum_{i=1}^n \Delta_iY_i/\sum_{i=1}^n \Delta_i)}{\sum_{i=1}^n \Delta_iX_i^2 - (\sum_{i=1}^n  \Delta_iX_i)^2/\sum_{i=1}^n\Delta_i}\right\} \\
&=& E_{X,\Delta}\left[\frac{\sum_{i=1}^n \Delta_iX_i\{E(Y_i|X_i,\Delta_i)-\sum_{i=1}^n \Delta_iE(Y_i|X_i,\Delta_i)/\sum_{i=1}^n \Delta_i\}}{\sum_{i=1}^n \Delta_iX_i^2 - (\sum_{i=1}^n  \Delta_iX_i)^2/\sum_{i=1}^n\Delta_i}\right].
\ese

Note that $E(Y_i|X_i,\Delta_i) = E(\epsilon_i|X_i,\Delta_i) + \beta_0 + \beta_1X_i =  E(\epsilon_i|\Delta_i) + \beta_0 + \beta_1X_i$ by our model assumptions in Section \ref{sec:notation}. Now, plugging back in to the numerator of the expectation, we have 
\bse
&& \sum_{i=1}^n \Delta_iX_i\left\{E(Y_i|X_i,\Delta_i)-\sum_{i=1}^n \Delta_iE(Y_i|X_i,\Delta_i)/\sum_{i=1}^n \Delta_i\right\} \\
&=& 
\sum_{i=1}^n \Delta_iX_i\left[E(\epsilon_i|\Delta_i) + \beta_0 + \beta_1X_i-\sum_{i=1}^n \Delta_i\left\{E(\epsilon_i|\Delta_i) + \beta_0 + \beta_1X_i \right\}/\sum_{i=1}^n \Delta_i\right].
\ese

Considering the terms with $\beta_0$, we have 
\bse
\sum_{i=1}^n \Delta_iX_i\left(\beta_0 - \sum_{i=1}^n \Delta_i\beta_0 / \sum_{i=1}^n \Delta_i\right) = \sum_{i=1}^n \Delta_iX_i(\beta_0 - \beta_0) = 0.
\ese

Considering the terms with $\beta_1$, we have
\bse
\sum_{i=1}^n \Delta_iX_i\left(\beta_1X_i - \sum_{i=1}^n \Delta_iX_i \beta_1 /\sum_{i=1}^n \Delta_i \right) = \beta_1 \left\{\sum_{i=1}^n\Delta_iX_i^2 - \left(\sum_{i=1}^n \Delta_iX_i\right)^2 / \sum_{i=1}^n \Delta_i\right\},
\ese
which will cancel out with the denominator from above leaving us with $\beta_1$. Finally, considering the constant terms (i.e., no parameters), we have 
\bse
\sum_{i=1}^n \Delta_iX_i \left\{ E(\epsilon_i|\Delta_i) - \sum_{i=1}^n \Delta_i E(\epsilon_i|\Delta_i)/\sum_{i=1}^n \Delta_i\right\}.
\ese 
If $\Delta_i = 0$, this whole term is $0$ due to the first $\Delta_i$ in the summation. If $\Delta_i = 1$, then $E(\epsilon_i|\Delta_i=1) = c$ where $c$ is a non-zero constant. Then the above expression is equal to 
\bse
\sum_{i=1}^n \Delta_iX_i \left( c - \sum_{i=1}^n \Delta_i c/\sum_{i=1}^n \Delta_i\right) = \sum_{i=1}^n \Delta_iX_i \left( c -  c \right) = 0.
\ese 
Now, we want to show that $\wh{\beta}_0$ may be biased when exogenous censoring doesn't hold. We show the expectation may not equal $\beta_0$ as follows: 
\bse
E(\wh{\beta}_0) &=& E_{X,\Delta}\left[E_{Y|X,\Delta}\left\{\sum_{i=1}^n \Delta_i(Y_i-\wh{\beta}_1X_i) / \sum_{i=1}^n \Delta_i\right\}\right] \\
&=&
E\left(1/ \sum_{i=1}^n \Delta_i\left[\left\{\sum_{i=1}^n \Delta_iE(Y_i|X_i,\Delta_i)\right\} - \sum_{i=1}^nE\left(\wh{\beta}_1|X_i,\Delta_i\right)\Delta_iX_i \right]\right) \\
&=&
E\left(1/\sum_{i=1}^n \Delta_i\left[\left\{\sum_{i=1}^n \Delta_iE(\epsilon_i|\Delta_i)+\Delta_i\beta_0+\Delta_i\beta_1X_i \right\} - \beta_1\sum_{i=1}^n\Delta_iX_i \right]\right) \\
&=&
\beta_0 + E(\epsilon|\Delta = 1),
\ese
which is not equal to $\beta_0$ unless $E(\epsilon|\Delta = 1) = 0$. If exogenous censoring does not hold, then $E(\epsilon|\Delta = 1) = E\{E(\epsilon|\Delta = 1, X)\}$ is not necessarily $0$, since $E(\epsilon|\Delta = 1, X) \neq 0$.

Finally, we consider what happens when there is no intercept, such that $m(\cdot) = \beta_1X$. The estimating equation from Section \ref{sec:bias_linear} will only have one component; namely, the second component from above, setting $\beta_0 = 0$. If we isolate $\beta_1$, we get $\wh{\beta}_1 = \sum_{i=1}^n \Delta_iX_iY_i / \sum_{i=1}^n \Delta_iX_i^2$. To investigate potential bias, we consider 
\bse
E(\wh{\beta}_1) &=& E\left(\sum_{i=1}^n \Delta_iX_iY_i / \sum_{i=1}^n \Delta_iX_i^2\right) \\
&=&
E\left\{\sum_{i=1}^n \Delta_iX_iE(Y_i|\Delta_i,X_i)/\sum_{i=1}^n \Delta_iX_i^2\right\} \\ 
&=&
E\left[\sum_{i=1}^n \Delta_iX_i\{E(\epsilon_i|\Delta_i)+\beta_1X_i\}/\sum_{i=1}^n \Delta_iX_i^2\right] \\ 
&=&
\beta_1 + E\left\{\sum_{i=1}^n \Delta_iX_iE(\epsilon_i|\Delta_i)/\sum_{i=1}^n \Delta_iX_i^2\right\},
\ese
which is equal to $\beta_1$ when exogenous censoring holds (i.e., $E(\epsilon_i|\Delta_i) = 0$), but may not be equal to $\beta_1$ otherwise.

\section{Justification for Simulation Settings}
\label{app:sims}

The following give justification to the data generation done for each simulation setting as shown in Table \ref{tab:sim_settings}. Each setting was chosen to satisfy (i) the respective combination of censoring mechanism assumptions and (ii) the chosen censoring rate, denoted as $r$. 

\begin{enumerate}[label=(\alph*.)]
    \item \textbf{Exogenous Censoring does not hold:} Mechanism \ref{assump:exogenous}\textendash\ref{assump:indep} do not hold.
    \begin{enumerate}[label=(\roman*.)]
        \item Mechanism \ref{assump:exogenous}, exogenous censoring, does not hold because $\bmu = E(\bepsilon|\bDelta, \bX, \bZ) \neq \bzero$ for both values of $\bDelta$. Specifically, $\bmu = \sigma/8$ when $\bDelta = 0$ and $\bmu = -\sigma/8 \times r/1-r$ when $\bDelta = 1$. 
    
    Even though exogenous censoring does not hold, the model errors must have conditional mean zero, i.e.,  $E(\bepsilon|\bX,\bZ) = \bzero$. To see that the errors have conditional mean zero, note that 
    \bse
    E(\bepsilon|\bX, \bZ) &=& E(\bepsilon|\bDelta = 1, \bX,\bZ)P(\bDelta=1|\bX,\bZ) \\
    &+& E(\bepsilon|\bDelta = 0, \bX,\bZ)P(\bDelta=0|\bX,\bZ) 
    \\
    &=&
    \frac{-\sigma}{8}\frac{r}{1-r}\times(1-r) + \frac{\sigma}{8}\times r = \bzero.
    \ese
    \item Since $\bDelta \sim \Bernoulli(1-r)$, this implies the probability of being observed is $1-r$, satisfying the desired censoring rate.
    \end{enumerate}
    
    \item \textbf{Strict Exogenous Censoring does not hold:} Mechanism \ref{assump:exogenous} holds and \ref{assump:strictexogenous}\textendash\ref{assump:indep}  do not hold. 
    \begin{enumerate}[label=(\roman*.)]
        \item Mechanism \ref{assump:strictexogenous}, strict exogenous censoring, does not hold by design since $\bDelta$ is directly generated dependent on $\bepsilon$ and independent of covariates $(\bX,\bZ)$. Therefore, the two are not independent given $(\bX, \bZ)$. This implies mechanisms \ref{assump:condxz}\textendash\ref{assump:indep}  also do not hold. However, exogenous censoring still holds because (1)  $\bepsilon$ is marginally generated from a normal distribution with mean $\bzero$, and (2) $\Delta$ is generated based on the \textit{absolute value} of $\epsilon$, so the expectation of $\epsilon$ given either value of $\Delta$ will still be $0$.
        \item  We define $\bDelta = 0$ when the absolute value of $\bepsilon$ is less than to the $r$th quantile of $|\bepsilon|$. The proportion of observations that are less than the $r$th quantile is exactly $r$, by definition of a quantile. This ensures that $r$ of the observations will have $\bDelta = 0$, satisfying the desired censoring rate.

    \end{enumerate}
    
    \item \textbf{Conditional Independence given (X,Z) does not hold:} Mechanism \ref{assump:exogenous}\textendash\ref{assump:strictexogenous} hold and \ref{assump:condxz}\textendash\ref{assump:indep}  do not hold. 
    \begin{enumerate}[label=(\roman*.)]
        \item In this setting, $\bepsilon$ is generated to be dependent on $\bC$, independent of covariates $(\bX, \bZ)$, so mechanisms \ref{assump:condxz}\textendash\ref{assump:indep}  do not hold. Particularly, $\bepsilon$ is more variable at the extremes of $\bC$. However, this says little about the relationship between $\bDelta$ and $\bepsilon$. If $\Delta_i =1$, then we know $X_i \leq C_i$, but even given $X_i$, this tells us nothing about whether or not $C_i \in \text{IQR}(C)$ and therefore tells us nothing about $\epsilon_i$. Therefore, $\bDelta$ is independent of $\bepsilon$ given $(\bX,\bZ)$, so censoring mechanisms \ref{assump:exogenous}\textendash\ref{assump:strictexogenous}  do hold.
        \item Now to see that we have generated the desired censoring rate, consider the following. Remember that $\bX \sim \Uniform(0,3)$. We consider $r < 0.5$ and $r >= 0.5$ separately, since for $r < 0.5$, $\max(C) > \max(X)$ and vice versa for $r > 0.5$. If $r < 0.5$, we define $\bC \sim \Uniform(0,3/(2r))$. Then
        \bse
        P(\bDelta = 0) &=& P(\bC \leq \bX) \\
        &=&
        \int_0^3\int_0^x \frac{2r}{3\times3} dc dx \\
        &=&
        \int_0^3 \frac{2r\times x}{9} dx \\
        &=& \frac{r}{9}(3^2-0^2) = r.
        \ese
        
        And if the censoring rate is $\geq 0.5$, then  $\bC \sim \Uniform(0,(6-6r))$ and
        \bse
        P(\bDelta = 1) &=& P(\bX \leq \bC) \\
        &=& \int_0^{6-6r}\int_0^c \frac{1}{3(6-6r)} dx dc \\
        &=& \int_0^{6-6r} \frac{c}{3(6-6r)} dc \\
        &=& \frac{1}{6(6-6r)}((6-6r)^2-0^2) = (6-6r)/6 = 1-r.
        \ese

    \end{enumerate}

    \item \textbf{Conditional Independence given Z does not hold:} Mechanism \ref{assump:exogenous}\textendash\ref{assump:condxz}  hold and \ref{assump:condz}\textendash\ref{assump:indep}  do not hold.
    \begin{enumerate}[label=(\roman*.)]
        \item $\bC$ is generated directly dependent on $\bX$, so Mechanism \ref{assump:condz}, conditionally independent censoring given $\bZ$, and therefore mechanism \ref{assump:indep}, independent censoring, cannot hold.      However, given $\bX$ and $\bZ$, $\bC$ is independent of $\bY$, so mechanism \ref{assump:condxz} and therefore \ref{assump:exogenous}\textendash\ref{assump:strictexogenous}  all hold. 
        \item To see that we have generated the desired censoring rate with $\bC \sim \Uniform(0, \bX/r)$,
        \bse
        P(\bDelta = 0) &=& P(\bC \leq \bX) \\
        &=& \int_0^3 \int_0^x \frac{r}{3x} dc dx \\
        &=& \int_0^3 \frac{r}{3} dx = \frac{r}{3}(3-0) = r.
        \ese
    \end{enumerate}

    \item \textbf{Independence does not hold:} Mechanism \ref{assump:exogenous}\textendash\ref{assump:condz}  hold and \ref{assump:indep}  does not hold. 
    \begin{enumerate}[label=(\roman*.)]
        \item $\bC$'s generation is directly dependent on $\bZ$, so mechanism \ref{assump:indep}, independent censoring, cannot hold. However $\bC$ is independent of $\bX$ and $\bY$ given $\bZ$, so mechanism \ref{assump:condz}  and therefore \ref{assump:exogenous}\textendash\ref{assump:condxz}  all hold. 
        \item To see that we have generated the desired censoring rate, we consider the generation of $\bC$. When $r \geq 0.5$, 50\% of the observations (those with $Z > \text{med}(\bZ)$ will be generated from $\bC \sim \Uniform(0,6-6r)$, which has censoring rate $r$ as shown in (c.ii.). The other 50\% (when $Z \leq \text{med}(\bZ)$) will be generated from $\bC \sim \Uniform(\frac{3-6r}{2-2r}, 3)$. This also has censoring rate $r$ as seen below:
        \bse 
         P(\bDelta = 0) &=& P(\bC \leq \bX) \\
         &=& \int_0^3 \int_{\frac{3-6r}{2-2r}}^3 \frac{1}{3\times(3-\frac{3-6r}{2-2r})} dc dx \\
         &=& \int_0^3 \frac{x-\frac{3-6r}{2-2r}}{3(3-\frac{3-6r}{2-2r})} dx \\
         &=& \frac{(3-\frac{3-6r}{2-2r})^2 - (\frac{3-6r}{2-2r})^2}{2\times3(3-\frac{3-6r}{2-2r})} \\
         &=& \frac{9-6\frac{3-6r}{2-2r}}{2\times3(3-\frac{3-6r}{2-2r})} \\
         &=& \frac{3-2\frac{3-6r}{2-2r}}{6-2\frac{3-6r}{2-2r}} \\
         &=& \frac{3(2-2r)-2(3-6r)}{6(2-2r)-2(3-6r)} = \frac{6r}{6} = r.
        \ese
    \end{enumerate}
Similarly, when $r < 0.5$, half of the observations will be generated from $\bC \sim \Uniform(0, \frac{3}{2r})$, which has censoring rate $r$ as shown in (c.ii). The other half will be generated from $\bC \sim \Uniform(3-6r, 3)$. This also has censoring rate $r$:
\bse
P(\bDelta = 1) &=& P(\bX \leq \bC) \\
&=& \int_{3-6r}^3\int_0^c \frac{1}{3(6r)} dx dc \\
&=& \int_{3-6r}^3 \frac{c}{3(6r)} dc \\
&=& \frac{3^2 -(3-6r)^2}{2*3(6r)} = \frac{-36r^2 +36r}{36r} = 1-r.
\ese
    \item \textbf{Independence holds:} Mechanisms (1-5) hold.
    \begin{enumerate}[label=(\roman*.)]
        \item  $\bC$ is generated independently of $\bX$, $\bY$, and $\bZ$, so mechanism \ref{assump:indep}  holds by design. This implies mechanisms \ref{assump:exogenous}\textendash\ref{assump:condz}  also hold.
        \item For justification how the choice of parameters satisfies the chosen censoring rate, see (c.ii.).

    \end{enumerate}

\end{enumerate}

\newpage

\section{Supplementary Tables}
\label{app:sim_tabs}
\renewcommand{\thetable}{S.\arabic{table}}

\begin{table}

\caption{\label{tab:sims_1200_linear} Simulation results for a linear model with a right-censored covariate when data are generated under six different combinations of censoring mechanism assumptions. We report the percent bias, estimated standard errors (SE), and 95\%  coverage probabilities for all parameters in the linear model when we estimate  parameters using the complete case estimator (CC) and oracle estimator (Oracle). Results are based on 1000 simulated datasets, each with a sample size of 1200.}
\centering
\resizebox{\linewidth}{!}{
\begin{tabular}[t]{llrrrrrrrrr}
\toprule
\multicolumn{2}{c}{\textbf{ }} & \multicolumn{3}{c}{\textbf{Percent Bias}} & \multicolumn{3}{c}{\textbf{Estimated SE}} & \multicolumn{3}{c}{\textbf{95\% Coverage}} \\
\cmidrule(l{5pt}r{5pt}){3-5} \cmidrule(l{5pt}r{5pt}){6-8} \cmidrule(l{5pt}r{5pt}){9-11}
\textbf{Censoring Rate} & \textbf{Method} & \textbf{$\wh{\beta}_0$} & \textbf{$\wh{\beta}_1$} & \textbf{$\wh{\beta}_2$} & \textbf{$\wh{\beta}_0$} & \textbf{$\wh{\beta}_1$} & \textbf{$\wh{\beta}_2$} & \textbf{$\wh{\beta}_0$} & \textbf{$\wh{\beta}_1$} & \textbf{$\wh{\beta}_2$}\\
\midrule
\addlinespace[0.3em]
\multicolumn{11}{c}{\textbf{Exogenous Censoring does not hold}}\\
 & CC & -11.14 & -0.21 & -0.10 & 0.09 & 0.05 & 0.05 & 0.91 & 0.94 & 0.94\\

\multirow[t]{-2}{*}{\raggedright\arraybackslash \hspace{1em}25\%} & Oracle & 0.45 & -0.08 & -0.10 & 0.08 & 0.05 & 0.04 & 0.93 & 0.92 & 0.94\\
\addlinespace
 & CC & -104.33 & -0.37 & -0.24 & 0.16 & 0.09 & 0.08 & 0.11 & 0.95 & 0.94\\

\multirow[t]{-2}{*}{\raggedright\arraybackslash \hspace{1em}75\%} & Oracle & 0.47 & -0.10 & -0.09 & 0.08 & 0.05 & 0.04 & 0.94 & 0.92 & 0.94\\
\addlinespace
\addlinespace[0.3em]
\multicolumn{11}{c}{\textbf{Strict Exogenous Censoring does not hold}}\\
 & CC & -1.19 & 0.29 & -0.06 & 0.11 & 0.06 & 0.05 & 0.96 & 0.95 & 0.95\\

\multirow[t]{-2}{*}{\raggedright\arraybackslash \hspace{1em}25\%} & Oracle & -0.95 & 0.22 & -0.04 & 0.08 & 0.05 & 0.04 & 0.95 & 0.94 & \vphantom{3} 0.95\\
\addlinespace
 & CC & -3.34 & 0.92 & 0.22 & 0.28 & 0.16 & 0.14 & 0.94 & 0.95 & 0.94\\

\multirow[t]{-2}{*}{\raggedright\arraybackslash \hspace{1em}75\%} & Oracle & -0.95 & 0.22 & -0.04 & 0.08 & 0.05 & 0.04 & 0.95 & 0.94 & \vphantom{3} 0.95\\
\addlinespace
\addlinespace[0.3em]
\multicolumn{11}{c}{\textbf{Conditional Independence given (X,Z) does not hold}}\\
 & CC & 0.63 & -0.18 & -0.14 & 0.09 & 0.05 & 0.05 & 0.95 & 0.94 & 0.95\\

\multirow[t]{-2}{*}{\raggedright\arraybackslash \hspace{1em}25\%} & Oracle & 0.53 & -0.08 & -0.11 & 0.08 & 0.05 & 0.04 & 0.93 & 0.93 & 0.95\\
\addlinespace
 & CC & 0.11 & 0.58 & 0.02 & 0.14 & 0.25 & 0.08 & 0.95 & 0.94 & 0.95\\

\multirow[t]{-2}{*}{\raggedright\arraybackslash \hspace{1em}75\%} & Oracle & 0.53 & -0.08 & -0.11 & 0.08 & 0.05 & 0.04 & 0.93 & 0.93 & 0.95\\
\addlinespace
\addlinespace[0.3em]
\multicolumn{11}{c}{\textbf{Conditional Independence given Z does not hold}}\\
 & CC & -0.95 & 0.22 & -0.06 & 0.09 & 0.05 & 0.05 & 0.96 & 0.94 & 0.95\\

\multirow[t]{-2}{*}{\raggedright\arraybackslash \hspace{1em}25\%} & Oracle & -0.95 & 0.22 & -0.04 & 0.08 & 0.05 & 0.04 & 0.95 & 0.94 & \vphantom{2} 0.95\\
\addlinespace
 & CC & -1.67 & 0.29 & 0.03 & 0.16 & 0.09 & 0.08 & 0.96 & 0.94 & 0.95\\

\multirow[t]{-2}{*}{\raggedright\arraybackslash \hspace{1em}75\%} & Oracle & -0.95 & 0.22 & -0.04 & 0.08 & 0.05 & 0.04 & 0.95 & 0.94 & \vphantom{2} 0.95\\
\addlinespace
\addlinespace[0.3em]
\multicolumn{11}{c}{\textbf{Independence does not hold}}\\
 & CC & -0.72 & 0.06 & -0.09 & 0.09 & 0.06 & 0.05 & 0.95 & 0.95 & 0.94\\

\multirow[t]{-2}{*}{\raggedright\arraybackslash \hspace{1em}25\%} & Oracle & -0.95 & 0.22 & -0.04 & 0.08 & 0.05 & 0.04 & 0.95 & 0.94 & \vphantom{1} 0.95\\
\addlinespace
 & CC & -1.40 & 0.62 & -0.20 & 0.13 & 0.14 & 0.09 & 0.94 & 0.93 & 0.93\\

\multirow[t]{-2}{*}{\raggedright\arraybackslash \hspace{1em}75\%} & Oracle & -0.95 & 0.22 & -0.04 & 0.08 & 0.05 & 0.04 & 0.95 & 0.94 & \vphantom{1} 0.95\\
\addlinespace
\addlinespace[0.3em]
\multicolumn{11}{c}{\textbf{Independence holds}}\\
 & CC & -1.00 & 0.23 & -0.06 & 0.09 & 0.06 & 0.05 & 0.95 & 0.94 & 0.95\\

\multirow[t]{-2}{*}{\raggedright\arraybackslash \hspace{1em}25\%} & Oracle & -0.95 & 0.22 & -0.04 & 0.08 & 0.05 & 0.04 & 0.95 & 0.94 & 0.95\\
\addlinespace
 & CC & -1.17 & 0.46 & -0.11 & 0.14 & 0.23 & 0.08 & 0.95 & 0.94 & 0.96\\

\multirow[t]{-2}{*}{\raggedright\arraybackslash \hspace{1em}75\%} & Oracle & -0.95 & 0.22 & -0.04 & 0.08 & 0.05 & 0.04 & 0.95 & 0.94 & 0.95\\
\bottomrule
\end{tabular}}
\end{table}

\begin{table}[t!]

\caption{\label{tab:sims_400_linear_int} 
Simulation results for a linear mean model with an intercept versus without an intercept.  We report the percent bias for all parameters in the linear model when we estimate parameters using the complete case estimator (CC) and oracle estimator (Oracle). Results are based on 1000 simulated datasets, each with a sample size of 400.}
\centering{
\begin{tabular}[t]{llrrr}
\toprule
\multicolumn{2}{c}{\textbf{ }} & \multicolumn{3}{c}{\textbf{Percent Bias}} \\
\cmidrule(l{5pt}r{5pt}){3-5}
\textbf{Censoring Rate} & \textbf{Method} & \textbf{$\wh{\beta}_0$} & \textbf{$\wh{\beta}_1$} & \textbf{$\wh{\beta}_2$}\\
\midrule
\addlinespace[0.3em]
\multicolumn{5}{c}{\textbf{Intercept Present}}\\
 & CC & -13.01 & 0.43 & 0.29\\

\multirow[t]{-2}{*}{\raggedright\arraybackslash \hspace{1em}0.25} & Oracle & -0.78 & 0.24 & 0.16\\
\addlinespace
 & CC & -109.01 & 0.72 & 0.38\\

\multirow[t]{-2}{*}{\raggedright\arraybackslash \hspace{1em}0.75} & Oracle & -0.67 & 0.20 & 0.17\\
\addlinespace
\addlinespace[0.3em]
\multicolumn{5}{c}{\textbf{No Intercept Present}}\\
 & CC & NA & -3.10 & -0.03\\

\multirow[t]{-2}{*}{\raggedright\arraybackslash \hspace{1em}0.25} & Oracle & NA & -0.13 & -0.01\\
\addlinespace
 & CC & NA & -26.76 & 0.21\\

\multirow[t]{-2}{*}{\raggedright\arraybackslash \hspace{1em}0.75} & Oracle & NA & -0.09 & -0.04\\
\bottomrule
\end{tabular}}
\end{table}

% logistic 400

\begin{table}

\caption{\label{tab:sims_400_logistic} 
Simulation results for a logistic model with a right-censored covariate when data are generated under six different combinations of censoring mechanism assumptions. We report the percent bias, estimated standard errors (SE), and 95\%  coverage probabilities for all parameters in the linear model when we estimate  parameters using the complete case estimator (CC) and oracle estimator (Oracle). Results are based on 1000 simulated datasets, each with a sample size of 400.}
\centering
\resizebox{\linewidth}{!}{
\begin{tabular}[t]{llrrrrrrrrr}
\toprule
\multicolumn{2}{c}{\textbf{ }} & \multicolumn{3}{c}{\textbf{Percent Bias}} & \multicolumn{3}{c}{\textbf{Estimated SE}} & \multicolumn{3}{c}{\textbf{95\% Coverage}} \\
\cmidrule(l{5pt}r{5pt}){3-5} \cmidrule(l{5pt}r{5pt}){6-8} \cmidrule(l{5pt}r{5pt}){9-11}
\textbf{Censoring Rate} & \textbf{Method} & \textbf{$\wh{\beta}_0$} & \textbf{$\wh{\beta}_1$} & \textbf{$\wh{\beta}_2$} & \textbf{$\wh{\beta}_0$} & \textbf{$\wh{\beta}_1$} & \textbf{$\wh{\beta}_2$} & \textbf{$\wh{\beta}_0$} & \textbf{$\wh{\beta}_1$} & \textbf{$\wh{\beta}_2$}\\
\midrule
\addlinespace[0.3em]
\multicolumn{11}{c}{\textbf{Exogenous Censoring does not hold}}\\
 & CC & -209.53 & 7.24 & 5.04 & 0.03 & 0.02 & 0.01 & 0.94 & 0.94 & 0.94\\

\multirow[t]{-2}{*}{\raggedright\arraybackslash \hspace{1em}25\%} & Oracle & -12.75 & 4.30 & 2.98 & 0.02 & 0.01 & 0.01 & 0.95 & 0.95 & 0.94\\
\addlinespace
 & CC & -1789.51 & 15.83 & 10.89 & 0.05 & 0.03 & 0.02 & 0.51 & 0.95 & 0.94\\

\multirow[t]{-2}{*}{\raggedright\arraybackslash \hspace{1em}75\%} & Oracle & -10.87 & 3.54 & 3.16 & 0.02 & 0.01 & 0.01 & 0.96 & 0.95 & 0.94\\
\addlinespace
\addlinespace[0.3em]
\multicolumn{11}{c}{\textbf{Strict Exogenous Censoring does not hold}}\\
 & CC & 5.73 & -0.71 & -1.07 & 0.03 & 0.02 & 0.02 & 0.94 & 0.94 & 0.94\\

\multirow[t]{-2}{*}{\raggedright\arraybackslash \hspace{1em}25\%} & Oracle & 2.65 & -0.39 & -0.93 & 0.02 & 0.01 & 0.01 & 0.94 & 0.95 & \vphantom{3} 0.94\\
\addlinespace
 & CC & -14.33 & 1.46 & -0.19 & 0.08 & 0.05 & 0.04 & 0.95 & 0.96 & 0.94\\

\multirow[t]{-2}{*}{\raggedright\arraybackslash \hspace{1em}75\%} & Oracle & 2.65 & -0.39 & -0.93 & 0.02 & 0.01 & 0.01 & 0.94 & 0.95 & \vphantom{3} 0.94\\
\addlinespace
\addlinespace[0.3em]
\multicolumn{11}{c}{\textbf{Conditional Independence given (X,Z) does not hold}}\\
 & CC & -22.46 & 7.63 & 3.41 & 0.02 & 0.02 & 0.01 & 0.96 & 0.96 & 0.94\\

\multirow[t]{-2}{*}{\raggedright\arraybackslash \hspace{1em}25\%} & Oracle & -11.37 & 3.35 & 2.21 & 0.02 & 0.01 & 0.01 & 0.95 & 0.94 & 0.94\\
\addlinespace
 & CC & -35.03 & 23.61 & 8.52 & 0.04 & 0.07 & 0.02 & 0.95 & 0.94 & 0.94\\

\multirow[t]{-2}{*}{\raggedright\arraybackslash \hspace{1em}75\%} & Oracle & -11.37 & 3.35 & 2.21 & 0.02 & 0.01 & 0.01 & 0.95 & 0.94 & 0.94\\
\addlinespace
\addlinespace[0.3em]
\multicolumn{11}{c}{\textbf{Conditional Independence given Z does not hold}}\\
 & CC & 1.92 & 2.62 & -1.06 & 0.03 & 0.02 & 0.01 & 0.94 & 0.94 & 0.93\\

\multirow[t]{-2}{*}{\raggedright\arraybackslash \hspace{1em}25\%} & Oracle & 2.65 & -0.39 & -0.93 & 0.02 & 0.01 & 0.01 & 0.94 & 0.95 & \vphantom{2} 0.94\\
\addlinespace
 & CC & -6.60 & 5.23 & -5.61 & 0.05 & 0.03 & 0.02 & 0.94 & 0.94 & 0.93\\

\multirow[t]{-2}{*}{\raggedright\arraybackslash \hspace{1em}75\%} & Oracle & 2.65 & -0.39 & -0.93 & 0.02 & 0.01 & 0.01 & 0.94 & 0.95 & \vphantom{2} 0.94\\
\addlinespace
\addlinespace[0.3em]
\multicolumn{11}{c}{\textbf{Independence does not hold}}\\
 & CC & 2.51 & 0.88 & -2.09 & 0.02 & 0.02 & 0.01 & 0.94 & 0.95 & 0.93\\

\multirow[t]{-2}{*}{\raggedright\arraybackslash \hspace{1em}25\%} & Oracle & 2.65 & -0.39 & -0.93 & 0.02 & 0.01 & 0.01 & 0.94 & 0.95 & \vphantom{1} 0.94\\
\addlinespace
 & CC & -8.00 & 13.40 & -5.54 & 0.04 & 0.04 & 0.02 & 0.93 & 0.93 & 0.93\\

\multirow[t]{-2}{*}{\raggedright\arraybackslash \hspace{1em}75\%} & Oracle & 2.65 & -0.39 & -0.93 & 0.02 & 0.01 & 0.01 & 0.94 & 0.95 & \vphantom{1} 0.94\\
\addlinespace
\addlinespace[0.3em]
\multicolumn{11}{c}{\textbf{Independence holds}}\\
 & CC & 2.34 & 0.35 & -2.49 & 0.02 & 0.02 & 0.01 & 0.94 & 0.95 & 0.93\\

\multirow[t]{-2}{*}{\raggedright\arraybackslash \hspace{1em}25\%} & Oracle & 2.65 & -0.39 & -0.93 & 0.02 & 0.01 & 0.01 & 0.94 & 0.95 & 0.94\\
\addlinespace
 & CC & 3.12 & 8.15 & -0.93 & 0.04 & 0.06 & 0.02 & 0.95 & 0.94 & 0.94\\

\multirow[t]{-2}{*}{\raggedright\arraybackslash \hspace{1em}75\%} & Oracle & 2.65 & -0.39 & -0.93 & 0.02 & 0.01 & 0.01 & 0.94 & 0.95 & 0.94\\
\bottomrule
\end{tabular}}
\end{table}

\begin{table}[t!]

\caption{\label{tab:sims_1200_logistic} Simulation results for a logistic model with a right-censored covariate when data are generated under six different combinations of censoring mechanism assumptions. We report the percent bias, estimated standard errors (SE), and 95\%  coverage probabilities for all parameters in the linear model when we estimate  parameters using the complete case estimator (CC) and oracle estimator (Oracle). Results are based on 1000 simulated datasets, each with a sample size of 1200.}
\centering
\resizebox{\linewidth}{!}{
\begin{tabular}[t]{llrrrrrrrrr}
\toprule
\multicolumn{2}{c}{\textbf{ }} & \multicolumn{3}{c}{\textbf{Percent Bias}} & \multicolumn{3}{c}{\textbf{Estimated SE}} & \multicolumn{3}{c}{\textbf{95\% Coverage}} \\
\cmidrule(l{5pt}r{5pt}){3-5} \cmidrule(l{5pt}r{5pt}){6-8} \cmidrule(l{5pt}r{5pt}){9-11}
\textbf{Censoring Rate} & \textbf{Method} & \textbf{$\wh{\beta}_0$} & \textbf{$\wh{\beta}_1$} & \textbf{$\wh{\beta}_2$} & \textbf{$\wh{\beta}_0$} & \textbf{$\wh{\beta}_1$} & \textbf{$\wh{\beta}_2$} & \textbf{$\wh{\beta}_0$} & \textbf{$\wh{\beta}_1$} & \textbf{$\wh{\beta}_2$}\\
\midrule
\addlinespace[0.3em]
\multicolumn{11}{c}{\textbf{Exogenous Censoring does not hold}}\\
 & CC & -178.68 & -3.33 & -1.63 & 0.02 & 0.01 & 0.01 & 0.91 & 0.94 & 0.94\\

\multirow[t]{-2}{*}{\raggedright\arraybackslash \hspace{1em}25\%} & Oracle & 7.18 & -1.19 & -1.47 & 0.01 & 0.01 & 0.01 & 0.93 & 0.92 & 0.94\\
\addlinespace
 & CC & -1697.33 & -3.60 & -1.10 & 0.03 & 0.02 & 0.01 & 0.12 & 0.95 & 0.95\\

\multirow[t]{-2}{*}{\raggedright\arraybackslash \hspace{1em}75\%} & Oracle & 7.51 & -1.45 & -1.28 & 0.01 & 0.01 & 0.01 & 0.94 & 0.92 & 0.94\\
\addlinespace
\addlinespace[0.3em]
\multicolumn{11}{c}{\textbf{Strict Exogenous Censoring does not hold}}\\
 & CC & -19.06 & 4.86 & -0.79 & 0.02 & 0.01 & 0.01 & 0.96 & 0.95 & 0.95\\

\multirow[t]{-2}{*}{\raggedright\arraybackslash \hspace{1em}25\%} & Oracle & -15.25 & 3.56 & -0.53 & 0.01 & 0.01 & 0.01 & 0.95 & 0.94 & \vphantom{3} 0.94\\
\addlinespace
 & CC & -53.55 & 16.10 & 5.29 & 0.05 & 0.03 & 0.02 & 0.95 & 0.95 & 0.95\\

\multirow[t]{-2}{*}{\raggedright\arraybackslash \hspace{1em}75\%} & Oracle & -15.25 & 3.56 & -0.53 & 0.01 & 0.01 & 0.01 & 0.95 & 0.94 & \vphantom{3} 0.94\\
\addlinespace
\addlinespace[0.3em]
\multicolumn{11}{c}{\textbf{Conditional Independence given (X,Z) does not hold}}\\
 & CC & 10.03 & -2.75 & -2.03 & 0.01 & 0.01 & 0.01 & 0.94 & 0.94 & 0.95\\

\multirow[t]{-2}{*}{\raggedright\arraybackslash \hspace{1em}25\%} & Oracle & 8.49 & -1.07 & -1.65 & 0.01 & 0.01 & 0.01 & 0.93 & 0.93 & 0.95\\
\addlinespace
 & CC & 2.11 & 9.86 & 0.89 & 0.02 & 0.04 & 0.01 & 0.95 & 0.94 & 0.95\\

\multirow[t]{-2}{*}{\raggedright\arraybackslash \hspace{1em}75\%} & Oracle & 8.49 & -1.07 & -1.65 & 0.01 & 0.01 & 0.01 & 0.93 & 0.93 & 0.95\\
\addlinespace
\addlinespace[0.3em]
\multicolumn{11}{c}{\textbf{Conditional Independence given Z does not hold}}\\
 & CC & -15.23 & 3.60 & -0.81 & 0.02 & 0.01 & 0.01 & 0.96 & 0.94 & 0.95\\

\multirow[t]{-2}{*}{\raggedright\arraybackslash \hspace{1em}25\%} & Oracle & -15.25 & 3.56 & -0.53 & 0.01 & 0.01 & 0.01 & 0.95 & 0.94 & \vphantom{2} 0.94\\
\addlinespace
 & CC & -26.86 & 5.14 & 1.08 & 0.03 & 0.02 & 0.01 & 0.96 & 0.94 & 0.95\\

\multirow[t]{-2}{*}{\raggedright\arraybackslash \hspace{1em}75\%} & Oracle & -15.25 & 3.56 & -0.53 & 0.01 & 0.01 & 0.01 & 0.95 & 0.94 & \vphantom{2} 0.94\\
\addlinespace
\addlinespace[0.3em]
\multicolumn{11}{c}{\textbf{Independence does not hold}}\\
 & CC & -11.66 & 1.05 & -1.31 & 0.01 & 0.01 & 0.01 & 0.95 & 0.95 & 0.94\\

\multirow[t]{-2}{*}{\raggedright\arraybackslash \hspace{1em}25\%} & Oracle & -15.25 & 3.56 & -0.53 & 0.01 & 0.01 & 0.01 & 0.95 & 0.94 & \vphantom{1} 0.94\\
\addlinespace
 & CC & -22.67 & 10.92 & -2.66 & 0.02 & 0.02 & 0.01 & 0.94 & 0.93 & 0.93\\

\multirow[t]{-2}{*}{\raggedright\arraybackslash \hspace{1em}75\%} & Oracle & -15.25 & 3.56 & -0.53 & 0.01 & 0.01 & 0.01 & 0.95 & 0.94 & \vphantom{1} 0.94\\
\addlinespace
\addlinespace[0.3em]
\multicolumn{11}{c}{\textbf{Independence holds}}\\
 & CC & -16.07 & 3.79 & -0.87 & 0.01 & 0.01 & 0.01 & 0.95 & 0.94 & 0.95\\

\multirow[t]{-2}{*}{\raggedright\arraybackslash \hspace{1em}25\%} & Oracle & -15.25 & 3.56 & -0.53 & 0.01 & 0.01 & 0.01 & 0.95 & 0.94 & 0.94\\
\addlinespace
 & CC & -19.06 & 8.16 & -1.28 & 0.02 & 0.04 & 0.01 & 0.95 & 0.95 & 0.96\\

\multirow[t]{-2}{*}{\raggedright\arraybackslash \hspace{1em}75\%} & Oracle & -15.25 & 3.56 & -0.53 & 0.01 & 0.01 & 0.01 & 0.95 & 0.94 & 0.94\\
\bottomrule
\end{tabular}}
\end{table}

\newpage

\begin{table}[t!]

\caption{\label{tab:hd_analysis_cap} 
Results from the PREDICT-HD study quantifying the relationship between apathy score and CAP score, controlling for other covariates. The goal of this analysis was to replicate the linear regression from \cite{Misiuraetal2019}. The parameters are estimated using the full dataset (n=781). We report the parameter estimate, the estimated standard error (SE), and the Wald-type 95\% confidence interval (95\% CI).}
\centering
\begin{tabular}[t]{rrrr}
\toprule
\textbf{Parameter} & \textbf{Estimate} & \textbf{Estimated SE} & \textbf{$95\%$ CI} \\
\midrule
\addlinespace[0.3em]

$\beta_0:$ Intercept & 12.83 & 0.29 & (12.25, 13.40) \\
$\beta_1:$ CAP Score & 0.23 & 0.50 & (-0.75, 1.23) \\
$\beta_2:$ Biological Sex & -0.49 & 0.33 & (-1.14, 0.16) \\
$\beta_3:$ Depression & 3.59 & 0.23 & (3.13, 4.04) \\
$\beta_4:$ Total Motor Score & 0.45 & 0.23 & (0.020, 0.89) \\
$\beta_5:$ Years of Education & 0.065 & 0.16 & (-0.26, 0.39) \\
$\beta_6:$ SDMT Score & -0.41 & 0.20 & (-0.80, -0.016) \\
$\beta_7:$ CAG Repeat Length & -0.079 & 0.52 & (-1.11, 0.95) \\
$\beta_8:$ Age & -0.43 & 0.45 & (-1.31, 0.44) \\

\bottomrule

\end{tabular}
\end{table}

\end{document}